\documentclass[a4paper,fleqn,usenatbib]{mnras}

\usepackage[T1]{fontenc}
\usepackage{ae,aecompl}

\usepackage{graphicx}	
\usepackage{amsmath}	
\usepackage{amssymb}	
\usepackage{cancel}
\usepackage{ulem}

\newcommand{\hMpc}{\,h^{-1}\, {\rm Mpc}}
\newcommand{\hGpc}{\,h^{-1}\, {\rm Gpc}}
\newcommand{\hNMpc}{\,h^{3}\, {\rm Mpc}^{-3}}
\newcommand{\hVGpc}{\,h^{-3}\, {\rm Gpc}^3}
\newcommand{\hk}{\,h\,{\rm Mpc^{-1}}}

\title[Limits on statistical anisotropy from BOSS DR12]{Limits on statistical anisotropy from BOSS DR12 galaxies using bipolar spherical harmonics}

\author[N. S. Sugiyama et al.]{
Naonori S. Sugiyama$^{1}$\thanks{E-mail: nao.s.sugiyama@gmail.com},
Maresuke Shiraishi$^{2,1}$,
and Teppei Okumura$^{3}$
\\
$^{1}$ Kavli Institute for the Physics and Mathematics of the Universe (WPI), \\
Todai Institutes for Advanced Study, The University of Tokyo, Chiba 277-8582, Japan\\
$^{2}$ Department of General Education, National Institute of Technology, Kagawa College, \\
355 Chokushi-cho, Takamatsu, Kagawa 761-8058, Japan \\
$^{3}$ Institute of Astronomy and Astrophysics, Academia Sinica, P. O. Box 23-141, Taipei 10617, Taiwan  \\
}

\date{}

\pubyear{2017}

\begin{document}
\label{firstpage}
\pagerange{\pageref{firstpage}--\pageref{lastpage}}
\maketitle

\begin{abstract}
We measure statistically anisotropic signatures imprinted in three-dimensional galaxy clustering using bipolar spherical harmonics (BipoSHs) in both Fourier space and configuration space. We then constrain a well-known quadrupolar anisotropy parameter $g_{2M}$ in the primordial power spectrum, parametrized by $P(\vec{k})  = \bar{P}(k) [ 1 + \sum_{M} g_{2M} Y_{2M}(\hat{k}) ]$, with $M$ determining the direction of the anisotropy. Such an anisotropic signal is easily contaminated by artificial asymmetries due to specific survey geometry. We precisely estimate the contaminated signal and finally subtract it from the data. Using the galaxy samples obtained by the Baryon Oscillation Spectroscopic Survey Data Release 12, we find no evidence for violation of statistical isotropy, $g_{2M}$ for all $M$ to be of zero within the $2\sigma$ level. The $g_{2M}$-type anisotropy can originate from the primordial curvature power spectrum involving a directional-dependent modulation $g_* (\hat{k} \cdot \hat{p})^2$. The bound on $g_{2M}$ is translated into $g_*$ as $-0.09 < g_* < 0.08$ with a $95\%$ confidence level when $\hat{p}$ is marginalized over.
  
\end{abstract}


\section{INTRODUCTION}
\label{Sec:Introduction}

The standard theory of inflation~\citep{Starobinsky:1980te,Sato:1980yn,Guth:1980zm,Linde:1981mu,Albrecht:1982wi} provides a highly successful mechanism for generating primordial density perturbations. The resulting perturbations are distributed as a statistically homogeneous, isotropic, parity-symmetric, and Gaussian random field. They provide the required seeds for the large-scale structure (LSS), after giving rise to temperature and polarization anisotropies in the cosmic microwave background (CMB) radiation, in excellent agreement with observation~\citep{Bennett:1996ce,Hinshaw2013ApJS..208...19H,Planck2016A&A...594A..13P,Eisenstein2005ApJ...633..560E,Alam:2016hwk}. Conversely, testing these fundamental properties is crucial in improving our understanding of the physics of the Universe and will provide us with hints for new physics.

The aim of this paper is to test a minimal deviation from the standard inflation model, a violation of statistical isotropy (SI) with preserving the other statistical properties of primordial fluctuations, using three-dimensional (3D) spectroscopic galaxy data of LSS surveys. We will focus especially on the so-called quadrupolar anisotropy~\citep{Ackerman2007PhRvD..75h3502A}, which is the simplest type of anisotropy that emerges from anisotropic inflation models in the limit of very weak anisotropy~(e.g., \citet{Dimopoulos2006PhRvD..74h3502D,Dimopoulos2008JHEP...07..119D,Yokoyama2008JCAP...08..005Y,Bartolo2009JCAP...10..015B,Bartolo2009JCAP...11..028B,Himmetoglu2009PhRvD..80l3530H,Himmetoglu2009PhRvL.102k1301H,Dimopoulos2010PhRvD..81b3522D,Gumrukcuoglu2010PhRvD..81f3528G,Watanabe2010PThPh.123.1041W,Soda2012CQGra..29h3001S,Bartolo2013PhRvD..87b3504B,Bartolo2015JCAP...01..027B,Bartolo2015JCAP...07..039B,Naruko2015JCAP...04..045N,Ashoorioon:2016lrg}) or an inflating solid or elastic medium \citep{Bartolo:2013msa, Bartolo:2014xfa}. The quadrupolar-type anisotropy is usually characterized by a parameter, $g_{2M}$ or $g_*$ (see Section \ref{Sec:Theory} and Section \ref{Sec:Results} for their definitions, respectively). Up to now, there are many constraints on the quadrupolar parameters, $g_{2M}$ and $g_*$, in CMB experiments~\citep{Groeneboom:2009fz,Groeneboom:2010cb,Hanson2009PhRvD..80f3004H,Bennett2011ApJS..192...17B,Bennett2013ApJS..208...20B,Kim2013PhRvD..88j1301K,Planck2016A&A...594A..16P,Planck2016A&A...594A..20P,Ramazanov:2016gjl} and in two-dimensional (2D) photometric catalogs of LSS~\citep{Pullen2010JCAP...05..027P}, yielding $|g_{2M}| \lesssim 10^{-2}$ ($10^{-1}$) from CMB (LSS). We should stress here that all of these observations are based on 2D analysis and test isotropy after projection along the line-of-sight (LOS). In contrast, a direct analysis on the 3D clustering will be free from the information loss due to the projection, achieving a more accurate test.

Our previous work (\citet{Shiraishi:2017wec}, hereafter S17) has proposed an application of bipolar spherical harmonics (BipoSHs;~\cite{Varshalovich1988qtam.book.....V,Hajian:2003qq,Hajian:2004zn,Hajian:2006ud}) to the redshift-space galaxy power spectrum and two-point correlation function treated in full 3D space. The power spectrum depending on two directions, wavevector $\vec{k}$ and the LOS direction $\hat{n}$, can be generally expanded in the BipoSH basis functions $\{Y_{\ell_1}(\hat{k})\otimes Y_{\ell_2}(\hat{n})\}_{LM}$, which are tensor products of spherical harmonics with two different arguments. The remarkable feature of the BipoSH formalism is to parameterize departures from SI regarding total angular momenta $L$ and $M$. In the absence of the assumption of SI, the corresponding expansion coefficients (hereafter, the BipoSH coefficients) yield the $L\geq1$ modes, while SI induces the $L=0$ mode alone. The BipoSH formalism can be therefore used to search for departures from SI and is always possible to translate any specific model for anisotropy. In S17, we have found that the $L=2$ mode is simply proportional to the quadrupolar parameter $g_{2M}$ in linear theory.

This work, for the first time, applies the BipoSH formalism to a publicly available 3D redshift survey data, the LOWZ and CMASS galaxy samples in both the North Galactic Cap (NGC) and the South Galactic Cap (SGC) derived from the Baryon Oscillation Spectroscopic Survey Data Release 12 (BOSS DR12;~\citet{Alam2015ApJS..219...12A}). We measure the $L=2$ mode of the BipoSH coefficients for both the power spectrum and the correlation function. We then constrain the anisotropy parameters, $g_{2M}$ and $g_*$, by comparing the measurements of the $L=2$ mode with their theoretical predictions developed in S17. Combining all the BOSS samples, CMASS and LOWZ, we find no evidence for violation of SI, namely $g_{2M}$ for all $M$ to be of zero within the $2\sigma$ level and $-0.09 < g_* < 0.08$ with $95\%$ probability, which is a more stringent constraint than the previous 2D analysis \citep{Pullen2010JCAP...05..027P}, as expected.

To reach the goal, we go through the following two steps. First, we develop an estimator of the BipoSH coefficients of both the power spectrum and the correlation function. Since the $L=0$ mode of the BipoSH coefficients reproduces the commonly used Legendre expansion coefficients (hereafter, the Legendre coefficients), our estimator can be regarded as a generalized one of the Legendre coefficients estimator~\citep{Landy1993ApJ...412...64L,Feldman1994ApJ...426...23F,Yamamoto2006PASJ...58...93Y,Bianchi2015MNRAS.453L..11B,Scoccimarro2015PhRvD..92h3532S}. Second, we discuss tools for studying the effects of survey geometry asymmetry on the observed BipoSH coefficients. Survey geometry asymmetries result in statistically anisotropic density fields. The asymmetries for their effects on the Legendre coefficients have already been studied in literature. We here confirm that they also produce ``mimic'' statistical anisotropies, biasing the primordial signal that we want to know. Estimating leakages of the anisotropic signal to the BipoSH coefficients, we find that the survey geometry effects provide a complete and sufficient explanation of the observed BipoSH coefficients, concluding a null detection of $g_{2M}$ or $g_*$.

The plan of our paper is as follows. In Section~\ref{Sec:Theory}, we briefly review the application of the BipoSH formalism to the galaxy power spectrum and correlation function discussed in S17. In Section~\ref{Sec:Data_Galaxy}, we summarize the galaxy sample data used in our analysis. In Section~\ref{Sec:Methodology}, we explain the technique to measure the BipoSH coefficients of the power spectrum and correlation function, followed by the treatment of the survey geometry effects in Section~\ref{Sec:SurveyWindowFunctions}. Section~\ref{Sec:Measurements} presents the measurements of the BipoSH coefficients and their covariance matrices. In Section~\ref{Sec:Results}, the results including constraints on the quadrupolar parameters, $g_{2M}$ and $g_*$, are presented. In this paper, we typically display figures only for CMASS NGC. However, we repeat the same analysis for the other three galaxy samples, CMASS SGC, LOWZ NGC, and  LOWZ SGC as that for CMASS NGC and constrain $g_{2M}$ and $g_*$ for all the four samples. We present a summary and conclusions in Section~\ref{Sec:Conclusions}. We additionally provide three Appendices: Appendix~\ref{Ap:Nneq0} presents the constraints on the quadrupolar parameter with various modulation scale-dependences, Appendix~\ref{Ap:Derivations} gives detailed derivations of equations used in our analysis, and Appendix~\ref{Ap:StandardDeviation} compares the standard deviation of the BipoSH coefficients estimated from mock catalogs with that computed by linear theory.

Throughout this paper we adopt a flat $\Lambda$CDM cosmology~\citep{Planck2016A&A...594A..13P}: $\Omega_{\rm m}=0.309$, $\Omega_{\rm \Lambda} = 0.691$, $n_{\rm s} = 0.9608$, and $H_0 = 100\, h\, {\rm km\, s^{-1}\, Mpc^{-1}}$ with $h=0.68$. We use the best fitting values of $f\sigma_8$ and $b\sigma_8$ measured in~\citet{Gil-Marn2016MNRAS.460.4188G}: $(f\sigma_8,\, b\sigma_8)=(0.392,\,1.283)$ for LOWZ and $(0.445,\,1.218)$ for CMASS.

\section{THEORY}
\label{Sec:Theory}

\subsection{BipoSH decomposition}
\label{Sec:biposh}

The theory of redshift space distortions (RSDs; see \cite{Hamilton1998ASSL..231..185H} for a review) is based on the redshift space to real space transformation,
\begin{eqnarray}
	\vec{x}  = \vec{x}_{\rm r} + \frac{\vec{v}(\vec{x}_{\rm r})\cdot\hat{x}_{\rm r}}{aH} \hat{x}_{\rm r},
\end{eqnarray}
where $\vec{x}$ is the three-dimensional coordinates of an observed galaxy, $\vec{x}_{\rm r}$ is the real-space position of the galaxy, $\hat{x}_{\rm r}=\vec{x}_{\rm r}/|\vec{x}_{\rm r}|$ is a unit vector pointing to the galaxy from the origin, $\vec{v}$ is the galaxy peculiar velocity, and $H$ is the Hubble expansion parameter. Under the global plane parallel approximation, the redshift-space galaxy power spectrum $P_{\rm g}$ is characterized by wavevector $\vec{k}$ and the LOS $\hat{n}$: $P_{\rm g}=P_{\rm g}(\vec{k},\hat{n})$. We note here that under the approximation, RSDs do preserve statistical homogeneity~\citep{Hamilton1998ASSL..231..185H}.

Any function depending on two directions can be expanded in spherical harmonics $Y_{\ell m}$ (e.g. \citet{Szapudi2004ApJ...614...51S}):
\begin{eqnarray}
	P_{\rm g}(\vec{k},\hat{n}) = \sum_{\ell m} \sum_{\ell'm'} P_{\ell m; \ell'm'}(k)\, y_{\ell m}(\hat{k})\, y_{\ell' m'}(\hat{n}),
\end{eqnarray}
where $y_{\ell m} = \sqrt{4\pi/(2\ell+1)}\, Y_{\ell m}$ is a normalized spherical harmonic function, and the corresponding expansion coefficients (hereafter, the spherical harmonic coefficients) $P_{\ell m;\ell' m'}(k)$ are given by
\begin{eqnarray}
	P_{\ell m; \ell' m'}(k) &=& (2\ell+1)(2\ell'+1) \int \frac{d^2\hat{k}}{4\pi}\int \frac{d^2\hat{n}}{4\pi} \nonumber \\
	&\times&  y^*_{\ell m}(\hat{k})\, y^*_{\ell' m'}(\hat{n})\, P_{\rm g}(\vec{k},\hat{n}).
	\label{Eq:Pllmm}
\end{eqnarray}

As an alternative way to generally decompose the power spectrum, we apply the BipoSH expansion~\citep{Varshalovich1988qtam.book.....V,Hajian:2003qq, Shiraishi:2017wec}:
\begin{eqnarray}
	  P_{\rm g}(\vec{k},\hat{n}) = \sum_{LM}\sum_{\ell\ell'} {\cal P}_{\ell\ell'}^{LM}(k)\, S_{\ell\ell'}^{LM}(\hat{k},\hat{n}).
\end{eqnarray}
In the above expression, we defined a normalized BipoSH basis $S_{\ell\ell'}^{LM}(\hat{k},\hat{n})$ as
\footnote{
The standard BipoSHs are given by
\begin{eqnarray}
	  Y_{\ell\ell'}^{LM}(\hat{n},\hat{n}') = \sum_{mm'}C_{\ell m; \ell' m'}^{LM} Y_{\ell m}(\hat{n})Y_{\ell' m'}(\hat{n}'),
\end{eqnarray}
where $C_{\ell m; \ell' m'}^{LM} = (-1)^{\ell-\ell'+M}\sqrt{2L+1} \left( \begin{smallmatrix} \ell & \ell' & L \\ m & m' & -M \end{smallmatrix}  \right)$ denote the Crebsh-Gordan coefficients, and this standard BipoSHs are related to our normalized ones as follows:
\begin{eqnarray}
	\scalebox{0.93}{$\displaystyle
	  Y_{\ell\ell'}^{LM}(\hat{n},\hat{n}') = (-1)^{\ell-\ell'} \sqrt{\frac{(2L+1)(2\ell+1)(2\ell'+1)}{(4\pi)^2}}  S_{\ell\ell'}^{LM}(\hat{n},\hat{n}').
$}
\end{eqnarray}
}
\begin{eqnarray}
	S_{\ell\ell'}^{LM}(\hat{k},\hat{n}) \equiv
	(-1)^{M}\sum_{mm'}
	\left( \begin{smallmatrix} \ell & \ell' & L \\ m & m' & -M \end{smallmatrix}  \right) y_{\ell m}(\hat{k})\, y_{\ell' m'}(\hat{n}),
	\label{Eq:BipoSH_basis}
\end{eqnarray}
where the matrices denote the Wigner 3-\textit{j} symbols, and the BipoSH coefficients ${\cal P}_{\ell\ell'}^{LM}$ are then given by
\begin{eqnarray}
	  {\cal P}_{\ell\ell'}^{LM}(k) = (2L+1) (-1)^{M} \sum_{mm'}
	  \left( \begin{smallmatrix} \ell & \ell' & L \\ m & m' & -M \end{smallmatrix}  \right) P_{\ell m;\ell' m'}(k).
	  \label{Eq:PLMll_from_Pllmm_0}
\end{eqnarray}
Throughout this paper, we use upper-case indices $LM$ for statistical anisotropies in the power spectrum and correlation function.

For $L=0$, the BipoSH coefficients ${\cal P}_{\ell\ell'}^{LM}$ are related to the Legendre coefficients:
\begin{eqnarray}
	  {\cal P}_{\ell\ell'}^{00}(k) = \delta_{\ell\ell'}\,\sqrt{2\ell+1}\, (-1)^{\ell}\, P_{\ell}(k),
\end{eqnarray}
where we used the relation $\left( \begin{smallmatrix} \ell & \ell' & 0 \\ m & m' & 0 \end{smallmatrix}  \right) = \frac{(-1)^{\ell'-m}}{\sqrt{2\ell+1}} \delta_{\ell\ell'}\delta_{m,-m'}$, and 
the Legendre coefficients $P_{\ell}$ are given by~\citep{Hamilton1998ASSL..231..185H}
\begin{eqnarray}
	  P_{\ell}(k) =  (2\ell+1) \int \frac{d^2\hat{k}}{4\pi}\int \frac{d^2\hat{n}}{4\pi}\, 
	  {\cal L}_{\ell}(\hat{k}\cdot\hat{n})\, P_{\rm g}(\vec{k},\hat{n})
\end{eqnarray}
with Legendre polynomials ${\cal L}_{\ell}$. In other words, the galaxy power spectrum can be written as
\begin{eqnarray}
	  P_{\rm g}(\vec{k},\hat{n})&=&\sum_{\ell} P_{\ell}(k)\, {\cal L}_{\ell}(\hat{k}\cdot\hat{n}) \nonumber \\
	  &+& \sum_{L\geq1,M}\sum_{\ell\ell'} {\cal P}_{\ell\ell'}^{LM}(k)\, S_{\ell\ell'}^{LM}(\hat{k},\hat{n}),
	  \label{Eq:Ledendre_and_BiPoSH}
\end{eqnarray}
where we used $S_{\ell\ell'}^{00}(\hat{k},\hat{n})=\delta_{\ell\ell'}\, ((-1)^{\ell}/\sqrt{2\ell+1})\,{\cal L}_{\ell}(\hat{k}\cdot\hat{n})$. If SI is valid, then the power spectrum can be only described by the Legendre coefficients. However, the presence of statistical anisotropy produces additional terms other than the Legendre coefficients. Rotational asymmetry terms $P_{\ell\ell'}^{L\geq1, M}$, i.e. non-zero total angular momenta, are orthogonal to the Legendre coefficients $P_{\ell}$ induced by the $L=0$ mode, which means that the $L\neq 0$ modes are unbiased observables of the rotational invariance breaking \citep{Shiraishi:2017wec}. 

For a practical analysis, S17 has defined a reduced BipoSH coefficients as
\begin{eqnarray}
	 P_{\ell\ell'}^{LM}(k) \equiv H_{\ell\ell'}^L\, {\cal P}_{\ell\ell'}^{LM}(k),
	  \label{Eq:PLMll_from_Pllmm}
\end{eqnarray}
where $H_{\ell\ell'}^L=\left( \begin{smallmatrix} \ell & \ell' & L \\ 0 & 0 & 0 \end{smallmatrix}  \right)$ filters even $\ell+\ell'+L$ components. In this paper, we only focus on the even $\ell+\ell'+L$ components, because a simple model breaking SI that we use in our analysis leads to ${\cal P}_{\ell\ell'}^{LM}\propto H_{\ell\ell'}^L$, i.e. $P_{\ell\ell'}^{LM}\propto (H_{\ell\ell'}^L)^2$ (see equation~(\ref{Eq:model_PLM})). We assume parity symmetry, i.e. invariance of the galaxy power spectrum under parity flip, $\vec{k}\to-\vec{k}$ and $\hat{n}\to-\hat{n}$, restricting allowed multipoles to $\ell+\ell'={\rm even}$. Therefore, our interests are only in the $L={\rm even}$ modes. Note that the filtering of $H_{\ell\ell'}^L$ provides a convenient normalization to reproduce the Legendre coefficients $P_{\ell}$ for $L=0$:
\begin{eqnarray}
	  P_{\ell\ell'}^{00}(k) = \delta_{\ell \ell'} P_{\ell}(k).
	  \label{Eq:PLMll_to_Pl}
\end{eqnarray}

The two-point correlation function can be expanded in spherical harmonics
\begin{eqnarray}
	  \xi(\vec{r},\hat{n}) = \sum_{\ell m}\sum_{\ell'm'} \xi_{\ell m;\ell' m'}(r)\,  y_{\ell m}(\hat{r})\, y_{\ell' m'}(\hat{n}),
	  \label{Eq:xi_SH}
\end{eqnarray}
and we define the reduced BipoSH coefficients as
\begin{eqnarray}
	\scalebox{0.91}{$\displaystyle
	  \xi_{\ell\ell'}^{LM}(r) \equiv (2L+1)\, H_{\ell\ell'}^L\, (-1)^{M} \sum_{mm'}
	  \left( \begin{smallmatrix} \ell & \ell' & L \\ m & m' & -M \end{smallmatrix}  \right) \xi_{\ell m;\ell' m'}(r).$}
	  \label{Eq:XiLMll_from_Pllmm}
\end{eqnarray}
Here, the spherical and BipoSH coefficients, $\xi_{\ell m;\ell' m'}$ and $\xi_{\ell\ell'}^{LM}$, are related to the Fourier-space ones according to the following Hankel transformations:
\begin{eqnarray}
	\xi_{\ell m;\ell' m'}(r) &=& i^{\ell}\int \frac{dk k^2}{2\pi^2}\, j_{\ell}(rk)\, P_{\ell m; \ell' m'}(k) \nonumber \\
	\xi_{\ell\ell'}^{LM}(r)  &=& i^{\ell}\int \frac{dk k^2}{2\pi^2}\, j_{\ell}(rk)\, P_{\ell \ell'}^{LM}(k),
	\label{Eq:P_LM_to_Xi_LM}
\end{eqnarray}
where $j_{\ell}$ is the spherical Bessel function of order $\ell$. Similarly to the power spectrum, the Legendre coefficients of the correlation function correspond to $\xi^{00}_{\ell\ell}$, given by
\begin{eqnarray}
	\xi_{\ell}(r) = \xi^{00}_{\ell \ell}(r) 
	= i^{\ell}\int \frac{dk k^2}{2\pi^2}\, j_{\ell}(rk)\, P_{\ell}(k).
	\label{Eq:xi_Legendre}
\end{eqnarray}

When the LOS direction is not determined by the global one $\hat{n}$ but by observed galaxy positions, the redshift-space power spectrum and correlation function become inhomogeneous, even if the primordial curvature perturbation $\zeta$ satisfies statistical homogeneity $\langle \zeta(\vec{k})\, \zeta(\vec{k}') \rangle = (2\pi)^3\,\delta_{\rm D}( \vec{k}+\vec{k}' )\,P_{\zeta}(k)$. This RSD-induced translational asymmetry significantly affects the observed power spectrum and correlation function through survey window functions, which is discussed in more detail in Section~\ref{Sec:SurveyWindowFunctions}.

\subsection{Quadrupolar-type anisotropy}
\label{Sec:QuadrupolarAnisotropy}

In linear theory, a galaxy power spectrum that breaks SI can be decomposed via~\citep{Shiraishi:2017wec}
\begin{equation}
	  P_{g}(\vec{k},\hat{n})=P_{\rm K}(\vec{k})
	  \left[ 1 + \sum_{L\geq 2}^{L = \text{even}} \sum_M g_{LM}\, f(k)\, Y_{LM}(\hat{k}) \right] ,
	  \label{Eq:parameterized_model}
\end{equation}
where $P_{\rm K}$ is the so-called Kaiser formula of linear RSD~\citep{Kaiser1987MNRAS.227....1K}, 
\begin{eqnarray}
	P_{\rm K}(\vec{k})=\left(\, \left(b\sigma_8 \right) + \left(f\sigma_8 \right)\, \mu_k^2\, \right)^2\, P_{\rm lin}(k,z=0),
\label{Eq:kaiser}
\end{eqnarray}
where $\mu_k=\hat{k}\cdot\hat{n}$, $\sigma_8$ denotes the rms matter fluctuation on scales of $8\hMpc$, $b$ represents the linear bias parameter, $f \sigma_8=d\ln \sigma_8/ \ln a$ is the logarithmic growth rate multiplied by $\sigma_8$, and $P_{\rm lin}$ is the isotropic, linear matter power spectrum. While $b$, $\sigma_8$, and $f$ are computed at given redshift $z$, $P_{\rm lin}$ is computed at $z=0$. In the above expression, $P_{\rm lin}$ is normalized by $\sigma_8^2$ so that $\sigma_8(z=0)=1$, because in the standard definition of $P_{\rm lin}$, it is proportional to $\sigma_8^2$. One can thus see from equation (\ref{Eq:kaiser}) that the amplitude of the observed power spectrum is characterized by the combination of $b\sigma_8$ and $f\sigma_8$, and that the RSD effect makes the LOS direction special.

The additional direction dependence is represented by $\sum_{LM} g_{LM} f(k) Y_{LM}(\hat{k})$, where $g_{LM}$ means the magnitude of statistical anisotropy on order $L$, with $M$ giving the direction of that anisotropy. The anisotropy parameter $g_{LM}$ satisfies a reality condition $g_{LM}^{\, *}(k) = (-1)^{M}g_{L,-M}$. The shape of the scale-dependence function of the primordial anisotropy, $f(k)$, depends strongly on the inflationary Lagrangian.\footnote{The linear growth rate $f$ should be distinguished from this $f(k)$.} In this analysis, we will treat $f(k)$ as a power law, $f(k)=(k/k_0)^n$, and consider four values of the spectral index, namely $n=-2,\,-1,\,0,\, {\rm and},\, 1$.\footnote{Such scale dependences are realized by, e.g., the running of an inflaton-vector coupling in vector inflation models \citep{Bartolo2013PhRvD..87b3504B,Bartolo2015JCAP...01..027B,Bartolo2015JCAP...07..039B}.} As the main results, we ignore the scale dependence of $f(k)$ and focus only on $n=0$, while we summarize the results for the other indexes, $n=-2,\,-1,\, {\rm and},\, 1$, in Appendix~\ref{Ap:Nneq0}.

Substituting the above equation~(\ref{Eq:parameterized_model}) into equation~(\ref{Eq:PLMll_from_Pllmm}) leads to
\begin{eqnarray}
	 P_{\ell\ell'}^{00}(k) &=& \delta_{\ell\ell'}\,P_{\ell'}(k) \nonumber \\
	 P_{\ell\ell'}^{L\geq2,M}(k)  &=& \sqrt{\frac{2L+1}{4\pi}}\, g_{LM}\, f(k)  \nonumber \\
	 &\times& (2\ell+1)\, (H^L_{\ell\ell'})^2\, P_{\ell'}(k),
	 \label{Eq:model_PLM}
\end{eqnarray}
where the Legendre coefficients $P_{\ell'}$ yield only the monopole $P_0  = \left( (b\sigma_8)^2 + (2/3)(b\sigma_8)(f\sigma_8) +(1/5)(f\sigma_8)^2 \right)P_{\rm lin}$, the quadrupole $P_2 = \left( (4/3) (b\sigma_8)(f\sigma_8) + (4/7) (f\sigma_8)^2 \right)P_{\rm lin}$, and the hexadecapole $P_4=(8/35)(f\sigma_8)^2P_{\rm lin}$ in linear theory, and these satisfy a magnitude relation $|P_0|>|P_2|>|P_4|$ at large scales. As expected, each of the $L\geq2$ modes is proportional to the corresponding anisotropy parameter $g_{LM}$, and the $L=0$ mode reproduces the Legendre coefficients $P_{\ell}$.

From now, we analyze the leading-order mode ($L=2$) in equation~(\ref{Eq:model_PLM}), the so-called ``quadrupolar anisotropy''. The parity-even condition and the triangular inequality coming from the filter $H_{\ell\ell'}^{L=2}$ restricts the allowed coefficients in $P_{\ell\ell'}^{LM}$ to $(\ell,\ell')=(2,0)$, $(0,2)$, $(2,2)$, $(4,2)$, $(2,4)$, $(4,4)$, and $(6,4)$. S17 has shown that $P_{20}^{2M}$ dominantly contributes to the signal-to-noise ratio, because $P_{20}^{2M}$ is proportional to the monopole $P_0$,  while all the other terms are proportional to the quadrupole $P_2$ or the hexadecapole $P_4$ that is smaller than the monopole.
Hence, we focus only on $P_{20}^{2M}$ in our analysis, which is derived from equation~(\ref{Eq:model_PLM}),
\begin{eqnarray}
	  P_{20}^{2M}(k) = \sqrt{\frac{5}{4\pi}}\, g_{2M}\, f(k)\, P_0(k).
	  \label{Eq:model_pk}
\end{eqnarray}
In the same manner as the Fourier-space analysis, in the configuration-space analysis we only consider $\xi_{20}^{2M}(r)$ given from equation~(\ref{Eq:P_LM_to_Xi_LM}) by
\begin{eqnarray}
	  \xi_{20}^{2M}(r) = - \sqrt{\frac{5}{4\pi}}\, g_{2M}\int \frac{dk k^2}{2\pi^2}\, j_2(rk)\, f(k)\, P_0(k),
	  \label{Eq:model_xi}
\end{eqnarray}
where we note here that $j_{\ell=2}$ is used, even though the power spectrum in the integrand is the monopole.

\section{DATA}
\label{Sec:Data_Galaxy}
We use two galaxy samples, the LOWZ sample with $463044$ galaxies between $z = 0.15\mathchar`-0.43$ $(z_{\rm eff} = 0.33)$ and the CMASS sample with $849637$ galaxies between $z = 0.43\mathchar`-0.7$ $(z_{\rm eff} = 0.56)$~\citep{White2011ApJ...728..126W,Parejko2013MNRAS.429...98P,Bundy2015ApJS..221...15B,Leauthaud2016MNRAS.457.4021L,Saito2016MNRAS.460.1457S}. These samples are drawn from the Data Release 12 (DR12; \citet{Alam2015ApJS..219...12A}) of the Baryon Oscillation Spectroscopic Survey (BOSS;~\citet{Bolton2012AJ....144..144B,Dawson2013AJ....145...10D}), which is part of the Sloan Digital Sky Survey III (SDSS-III;~\citet{Eisenstein2011}), and are selected from multi-color SDSS imaging~\citep{Fukugita1996,Gunn1998,Smith2002AJ....123.2121S,Gunn2006,Doi2010AJ....139.1628D}. 

To correct for several observational artifacts in the catalogs and obtain unbiased estimates of the galaxy density field, we use a completeness weight for each galaxy~\citep{Ross2012MNRAS.424..564R,Anderson2014MNRAS.441...24A,Reid2016MNRAS.455.1553R},
\begin{eqnarray}
	  w_{ {\rm c}}(\vec{x}\,) = w_{ {\rm systot}}(\vec{x}\,) \left( w_{ {\rm cp}}(\vec{x}\,) + w_{ {\rm noz}}(\vec{x}\,) - 1\right),
	  \label{Eq:weight_c}
\end{eqnarray}
where $\vec{x}$ is the observed galaxy position, and $w_{\rm cp}$, $w_{\rm noz}$, and $w_{\rm systot}$ denote a redshift failure weight, a collision weight, and a angular systematics weight, respectively. The details about the observational systematic weights are described in~\citet{Reid2016MNRAS.455.1553R}. Additionally, we use the optimal weighting of galaxies, so-called the FKP weight $w_{\rm FKP}$~\citep{Feldman1994ApJ...426...23F}. We adopt the values of the FKP weight given in the publicly available DR12 galaxy and random catalogues, which are computed using the following amplitude of the power spectrum, $P_0 = 10^{4}\, h^{-3}\, {\rm Mpc}^{3}$. However, we do not expect that more appropriate values of $P_0$ will significantly improve the constraint on the anisotropy. By multiplying the completeness weight by the FKP weight, we finally define a local weight function that we use in our analysis: 
\begin{eqnarray}
	  w(\vec{x}\,) = w_{\rm c}(\vec{x}\,)\, w_{\rm FKP}(\vec{x}\,).
	  \label{Eq:weight}
\end{eqnarray}

\section{METHODOLOGY}
\label{Sec:Methodology}

In this section, we describe the estimators we use to measure the power spectrum and correlation function from the observed galaxy distribution.

The number density field of galaxies is given by
\begin{eqnarray}
	  n(\vec{x}\,) = \sum_i^{N_{\rm gal}}\, w(\vec{x}_i)\, \delta_{\rm D}\left( \vec{x} - \vec{x}_i \right),
	  \label{Eq:density_gal}
\end{eqnarray}
where $\vec{x}_i$ represents the observed position of galaxy $i$, the weight function $w(\vec{x}\,)$ is given by equation~(\ref{Eq:weight}), $N_{\rm gal}$ denotes the total number of observed galaxies, and $\delta_{\rm D}$ is a Dirac $\delta$-function. To estimate the galaxy density fluctuation $\delta n$, we measure the mean number density $\bar{n}(\vec{x}\,)$ from a synthetic random catalog\footnote{For the random catalogs of the LOWZ and CMASS samples, we do not need the completeness weight in equation~(\ref{Eq:weight_c}): $w_{\rm c}=1$.}, multiplied by a factor $\alpha$,
\begin{eqnarray}
	  \bar{n}(\vec{x}\,) = \alpha\, \sum_i^{N_{\rm ran}}\, w(\vec{x}_i)\, \delta_{\rm D}\left( \vec{x} - \vec{x}_i\right),
	  \label{Eq:density_ran}
\end{eqnarray}
where $N_{\rm ran}$ represents the total number of objects in the random catalog, and $\alpha$ is the ratio between the weighted numbers of galaxies in the real and random catalogs: $\alpha =  \sum_i^{N_{\rm gal}}w(\vec{x}_i) / \sum_{i}^{N_{\rm ran}}w(\vec{x}_i) \sim 0.01$ in our analysis. By subtracting $\bar{n}(\vec{x}\,)$ from $n(\vec{x}\,)$, we obtain the observed galaxy density fluctuation
\begin{eqnarray}
	\delta n(\vec{x}\,) = n(\vec{x}\,) - \bar{n}(\vec{x}\,).
	\label{Eq:fluctuation}
\end{eqnarray}

\subsection{Power spectrum}
\label{Sec:PowerSpectrum}

In analogy to the estimator of the Legendre coefficients of the power spectrum~\citep{Feldman1994ApJ...426...23F,Yamamoto2006PASJ...58...93Y}, we present an estimator of $P_{\ell m;\ell' m'}$ (equation~\ref{Eq:Pllmm}) as follows
\begin{eqnarray}
	\scalebox{0.95}{$\displaystyle
	\widehat{P}_{\ell m;\ell' m'}(k) $}
	\hspace{-0.25cm}
	 &=& 
	\hspace{-0.25cm}
	\scalebox{0.95}{$\displaystyle
	\frac{(2\ell+1)(2\ell'+1)}{A} \int \frac{d^2\hat{k}}{4\pi}\, y_{\ell m}^{*}(\hat{k}) \int d^3x_1\int d^3x_2 $} \nonumber \\
	&\times&
	\hspace{-0.25cm}
	\scalebox{0.95}{$\displaystyle
	 e^{-i\vec{k}\cdot\vec{x}_{12}}\,  y_{\ell' m'}^{*}(\hat{n}_{12})\, \delta n(\vec{x}_1)\, \delta n(\vec{x}_2), $}
	\label{Eq:estimator_Pllmm_natural}
\end{eqnarray}
where $\vec{x}_{12} = \vec{x}_1 - \vec{x}_2$ is the relative coordinates of the pair of points $\vec{x}_1$ and $\vec{x}_2$, the unit vector of $\vec{n}_{12}=(\vec{x}_1+\vec{x}_2)/2$, denoted as $\hat{n}_{12}$, is used as the LOS direction to the pair, and $A$ is the normalization factor given by
\begin{eqnarray}
	  A = \int d^3x\, \bar{n}^2(\vec{x}\,).
	  \label{Eq:normalization}
\end{eqnarray}
This normalization depends on a grid-cell resolution to compute the density field, and therefore, it is difficult to make $A$ converge to a certain value in a large survey. However, this difficulty does not necessarily become an issue, because the value of $A$ does not affect the final results as we will see in Section~\ref{Sec:SurveyWindowFunctions}. The integral $\int d^2\hat{k}$ in equation (\ref{Eq:estimator_Pllmm_natural}) is the angular integration, and it is performed over a spherical shell in Fourier space centered at each bin $k=|\vec{k}|$,
\begin{eqnarray}
	  \int \frac{d^2\hat{k}}{4\pi} = \frac{1}{N_k} \sum_{k-\Delta k/2 < k < k + \Delta k /2},
\end{eqnarray}
where $\Delta k$ is the bin size, and $N_k = 4\pi k^2 \Delta k V / (2\pi)^3$ is the number of independent Fourier modes with $V$ being a given survey volume. Finally, equation~(\ref{Eq:PLMll_from_Pllmm}) relates the estimator of $P_{\ell m;\ell'm'}$ to that of $P_{\ell\ell'}^{LM}$ through
\begin{eqnarray}
	  \widehat{P}_{\ell\ell'}^{\,LM}(k) &\equiv& (2L+1)\, H_{\ell\ell'}^L\, (-1)^{M} \nonumber \\
	  &\times& 
	 \sum_{mm'}\, \left( \begin{smallmatrix} \ell & \ell' & L \\ m & m' & -M \end{smallmatrix}  \right)\, \widehat{P}_{\ell m;\ell' m'}(k).
	  \label{Eq:estimator_PLM}
\end{eqnarray}
In particular, for $\widehat{P}_{20}^{2M}$ we obtain
\begin{eqnarray}
	\widehat{P}_{20}^{\,2M}(k) &=& \widehat{P}_{2M;00}(k) \nonumber \\
	&=& 
	\frac{5}{A} \int \frac{d^2\hat{k}}{4\pi}\, y_{2M}^{*}(\hat{k}) \left|\delta n(\vec{k})\right|^2,
\end{eqnarray}
where $\delta n(\vec{k})$ is the Fourier transform of $\delta n(\vec{x})$, given by
\begin{eqnarray}
	  \delta n(\vec{k}) = \int d^3x\, e^{-i\vec{k}\cdot\vec{x}}\, \delta n(\vec{x}).
\end{eqnarray}

To compute $\widehat{P}_{\ell m;\ell' m'}$ and $\widehat{P}_{\ell\ell'}^{\, LM}$, we apply the local plane parallel approximation $\hat{x}_1\approx\hat{x}_2$, which is known to be rather accurate for the Legendre coefficients of the power spectrum $P_{\ell}$~\citep{Samushia2015MNRAS.452.3704S}. This approximation allows the integrals in equation~(\ref{Eq:estimator_Pllmm_natural}) to decouple into a product of Fourier transforms:
\begin{eqnarray}
	\widehat{P}_{\ell m;\ell' m'}(k) &=&
	\frac{(2\ell+1)(2\ell'+1)}{A} \int \frac{d^2\hat{k}}{4\pi}\,  y_{\ell m}^{*}(\hat{k}) \nonumber \\
	&\times& \delta n_{\ell' m'}(\vec{k})\, \delta n^*(\vec{k}), 
	\label{Eq:estimator_Pllmm_fft}
\end{eqnarray}
where $\delta n_{\ell m}(\vec{k})$ is given by
\begin{eqnarray}
	  \delta n_{\ell m}(\vec{k}\,) = \int d^3x\, e^{-i\vec{k}\cdot\vec{x}}\, y_{\ell m}^{*}(\hat{x})\, \delta n(\vec{x}\,).
\end{eqnarray}
The density fluctuation $\delta n(\vec{x}\,)$ multiplied by $y^*_{\ell m}(\hat{x})$, which is denoted as $\delta n_{\ell m}(\vec{x}\,)$, can be directly measured from a galaxy sample as,
\begin{eqnarray}
	\hspace{-0.5cm}\delta n_{\ell m}(\vec{x}\,) \hspace{-0.25cm}&=& \hspace{-0.25cm} y_{\ell m}^{*}(\hat{x})\, \delta n(\vec{x}\,) \nonumber \\
	  &=&\hspace{-0.25cm}
	  \left( \sum_i^{N_{\rm gal}} - \alpha \sum_i^{N_{\rm ran}}  \right)
	  y_{\ell m}^{*}(\hat{x}_i)\,  w(\vec{x}_i)\, \delta_{\rm D}\left( \vec{x}-\vec{x}_i \right).
\end{eqnarray}
Therefore, we stress here that $\delta n_{\ell m}(\vec{k})$ is computable using any fast Fourier transform (FFT) algorithm as the Fourier transform of $\delta n_{\ell m}(\vec{x})$. The computation of equation~(\ref{Eq:estimator_Pllmm_fft}) then will be of ${\cal O}(N_k \ln N_k)$.

The FFT algorithm requires the interpolation of functions on a regular grid in position space. The Fourier transform of the density fluctuation measured by FFTs, $\delta n_{\ell m}(\vec{k}\,)|_{\rm FFT}$, includes the effect of the mass assignment function $W_{\rm mass}(\vec{k}\,)$~\citep{Jing2005ApJ...620..559J}. We can remove such effects from $\delta n_{\ell m}(\vec{k})|_{\rm FFT}$ by simply dividing by $W_{\rm mass}(\vec{k})$: $\delta n_{\ell m}(\vec{k}) = \delta n_{\ell m}(\vec{k})|_{\rm FFT}/W_{\rm mass}(\vec{k})$. The most popular mass assignment function is given by
\begin{eqnarray}
	  W_{\rm mass}(\vec{k}) = \prod_{i=x,y,z}\left[ {\rm sinc}\left( \frac{\pi k_i}{2 k_{{\rm N},i}} \right) \right]^{p},
\end{eqnarray}
where $k_{{\rm N},i}=\pi/H_i$ is the Nyquist frequency of $i$-axis with the grid spacing $H_i$ on the axis. The indexes $p=1$, $p=2$, and $p=3$ correspond to the nearest grid point (NGP), cloud-in-cell (CIC), and triangular-shaped cloud (TSC) assignment functions, respectively.

Finally, we need to subtract shot-noise terms from the BipoSH coefficients $\widehat{P}_{\ell\ell'}^{\, LM}$ computed by equation~(\ref{Eq:estimator_PLM}). The shot-noise terms $S_{\ell \ell'}^{LM}$ on the BipoSH coefficients are given by
\begin{eqnarray}
	  S_{\ell\ell'}^{LM}(k)
	\hspace{-0.25cm}
	  &=& 
	\hspace{-0.25cm}
	  (2L+1)\, H_{\ell\ell'}^L (-1)^M \nonumber \\
	  &\times& \sum_{mm'} \left( \begin{smallmatrix} \ell & \ell' & L \\ m & m' & -M \end{smallmatrix}  \right) S_{\ell m;\ell' m'}(k)
\end{eqnarray}
with
\begin{eqnarray}
	\hspace{-0.3cm} S_{\ell m;\ell' m'}(k) 
	\hspace{-0.25cm}
	&=&
	\hspace{-0.25cm}
	\frac{(2\ell+1)(2\ell'+1)}{A}
	  \int \frac{d^2\hat{k}}{4\pi}  y_{\ell m}^{*}(\hat{k}) \frac{C_{\rm shot}(\vec{k})}{W_{\rm mass}^2(\vec{k})}
	  \nonumber \\
	  &\times&  \left( \sum_{i}^{N_{\rm gal}}+ \alpha^2 \sum_{i}^{N_{\rm ran}} \right) [w(\vec{x}_i\,)]^2\, y_{\ell' m'}^{ *}(\vec{x}_i).
\end{eqnarray}
Here, the function $C_{\rm shot}(\vec{k})$ has a simple analytic function given by equation (20) in~\cite{Jing2005ApJ...620..559J}. 

The $L=0$ mode of the BipoSH estimator $\widehat{P}_{\ell\ell'}^{LM}$ reproduces the Legendre coefficients estimator from equation~(\ref{Eq:PLMll_to_Pl}),
\begin{eqnarray}
	  \widehat{P}_{\ell}(k) = \widehat{P}_{\ell\ell}^{00}(k)
	  = \frac{1}{2\ell+1}\sum_{m=-\ell}^{\ell}(-1)^{m}\widehat{P}_{\ell, -m; \ell m}(k), \label{eq:estimator_pl}
\end{eqnarray}
where we used the relation ${\cal L}_{\ell}(\hat{x}\cdot\hat{y})= \sum_{m} y_{\ell m}(\hat{x}) y^*_{\ell m}(\hat{y})$. Therefore, our BipoSH estimator can be used as an alternative to the standard FFT-based method to measure the multipole moments $P_\ell$~\citep{Bianchi2015MNRAS.453L..11B,Scoccimarro2015PhRvD..92h3532S}. However, we stress here that one can measure $P_\ell$ with $\ell > 0$ faster using our estimator with the $L=0$ mode (equation~\ref{eq:estimator_pl}) than the standard one, because the BipoSH estimator requires a smaller number of FFTs to compute $\widehat{P}_{\ell}$: $\widehat{P}_\ell$ can be measured by our estimator by $(2\ell+1)$ FFTs, while the standard one requires $1$, $6$ and $15$ FFTs for $\ell=0$, $2$ and $4$, repectively.

\subsection{Two-point correlation function}
\label{Sec:TwoPointCorrelationFunction}
Now we move onto the derivation of the estimator for the BipoSH coefficients of the two-point correlation function. To clarify the relation between the estimators of the power spectrum and the correlation function, we first present the estimator for the coefficients normalized by the factor $A$ (equation~\ref{Eq:normalization}), $\xi_{\ell \ell'}^{LM}|_{\rm A}$:
\begin{eqnarray}
	\widehat{\xi}_{\ell \ell'}^{\, LM}(r)\big|_{\rm A}
	&=& 
	 (2L+1)\, H_{\ell\ell'}^L (-1)^M \nonumber \\
	  &\times& \sum_{mm'} \left( \begin{smallmatrix} \ell & \ell' & L \\ m & m' & -M \end{smallmatrix}  \right) 
	\widehat{\xi}_{\ell m;\ell' m'}(r)\big|_{\rm A},
	\label{Eq:estimator_XillLM_A}
\end{eqnarray}
where
\begin{eqnarray}
	\scalebox{0.95}{$\displaystyle
		\widehat{\xi}_{\ell m;\ell' m'}(r)\big|_{\rm A} $}
	\hspace{-0.25cm}
	 &=& 
	\hspace{-0.25cm}
	\scalebox{0.95}{$\displaystyle
	\frac{(2\ell+1)(2\ell'+1)}{A} \int \frac{d^2\hat{r}}{4\pi}\, y_{\ell m}^{*}(\hat{r}) 
	\int d^3x_1\int d^3x_2 $} \nonumber \\
	&\times&
	\scalebox{0.95}{$\displaystyle
		\delta_{\rm D}\left( \vec{r} - \vec{x}_{12}\right)\,  y_{\ell' m'}^{*}(\hat{n}_{12})\, \delta n(\vec{x}_1)\, \delta n(\vec{x}_2) $}.
	\label{Eq:estimator_Xillmm_A}
\end{eqnarray}
A Hankel transform relates the above estimator $\widehat{\xi}_{\ell \ell'}^{\, LM}|_{\rm A}$ to $\widehat{P}_{\ell\ell'}^{\, LM}$ (equation~\ref{Eq:estimator_PLM}) as,
\begin{eqnarray}
	\widehat{P}_{\ell \ell' }^{\, LM}(k) = 4\pi(-i)^{\ell} \int dr\, r^2\, j_{\ell}(kr)\, \widehat{\xi}_{\ell \ell'}^{\, LM}(r)\big|_{\rm A}.
	\label{Eq:XI_TO_PK}
\end{eqnarray}
To compute $\widehat{\xi}_{\ell \ell'}^{\, LM}|_{\rm A}$, we propose two ways, the pair-counting approach (e.g., \citet{Landy1993ApJ...412...64L}) and the FFT-based approach (e.g., \citet{Slepian:2015qwa}). First, substituting equations~(\ref{Eq:density_gal}) and (\ref{Eq:density_ran}) into equation~(\ref{Eq:estimator_Xillmm_A}) leads to the pair-counting estimator, reading
\begin{eqnarray}
	\scalebox{0.85}{$\displaystyle
	\widehat{\xi}_{\ell m;\ell'm'}(r)\big|_{\rm A} $}
	\hspace{-0.25cm}
	 &=& 
	\hspace{-0.25cm}
	\scalebox{0.85}{$\displaystyle
	 \frac{(2\ell+1)(2\ell'+1)}{4\pi r^2 \Delta r A }	
	\left(  \sum_{i,j}^{N_{\rm gal}} - 2 \alpha \sum_i^{N_{\rm gal}} \sum_j^{N_{\rm ran}} 
	+ \alpha^2 \sum_{i,j}^{N_{\rm ran}}  \right) $} \nonumber \\
	&\times&
	\hspace{-0.25cm}
	\scalebox{0.85}{$\displaystyle
	\delta_{\rm K}\left( r - |\vec{x}_{ij}|\right)\,  y_{\ell m}^{*}(\hat{x}_{ij})\,  y_{\ell' m'}^{*}(\hat{n}_{ij})\, w(\vec{x}_i)\,w(\vec{x}_j), $} 
	\label{Eq:estimator_Xillmm_pair}
\end{eqnarray}
where $\hat{n}_{ij}$ and $\hat{x}_{ij}$ are respectively the unit vectors of $\vec{n}_{ij} = (\vec{x}_i+\vec{x}_j)/2$ and $\vec{x}_{ij} = \vec{x}_i - \vec{x}_j$, $\delta_{\rm K}$ is the Kronecker delta, and $\Delta r$ is the bin size. The snot-noise term, which only contributes to a bin of $r=0$, can be removed by not counting $i=j$ from the summations in equation~(\ref{Eq:estimator_Xillmm_pair}), $\sum_{i,j}^{N_{\rm gal}}$ and $\sum_{i,j}^{N_{\rm ran}}$. Second, under the local plane parallel approximation $\hat{x}_1\approx\hat{x}_2$, the correlation function can be computed by FFTs
\begin{eqnarray}
	\hspace{-0.15cm}
	\scalebox{0.95}{$\displaystyle
	\widehat{\xi}_{\ell m;\ell' m'}(r) \big|_{\rm A} $}
	\hspace{-0.25cm}
	&=& 
	\hspace{-0.25cm}
	\scalebox{0.95}{$\displaystyle
	\frac{(2\ell+1)(2\ell'+1)}{A} \int \frac{d^2\hat{r}}{4\pi}\, y_{\ell m}^{*}(\hat{r}) 
	\int \frac{d^3k}{(2\pi)^3}\, e^{i\vec{k}\cdot\vec{r}}
	$} \nonumber \\
	&\times&
	\hspace{-0.25cm}
	\scalebox{0.95}{$\displaystyle
	\left[  \delta n_{\ell' m'}(\vec{k})\, \delta n^*(\vec{k}) - S_{\ell' m'}(\vec{k}) \right], $}
	\label{Eq:estimator_Xillmm_fft}
\end{eqnarray}
where the shot-noise terms $S_{\ell m}(\vec{k})$ are given by
\begin{eqnarray}
	\scalebox{0.95}{$\displaystyle
	  S_{\ell m}(\vec{k}) = 
	  \frac{C_{\rm shot}(\vec{k})}{W_{\rm mass}^2(\vec{k})}
	  \left( \sum_{i}^{N_{\rm gal}}+ \alpha^2 \sum_{i}^{N_{\rm ran}} \right) [w(\vec{x}_i\,)]^2\, y_{\ell m}^{*}(\vec{x}_i). $}
\end{eqnarray}
The pair-counting approach has an advantage in a robust estimation of the correlation function at small scales compared to the FFT-based approach. On the other hand, the FFT-based approach is faster than pair-counting algorithms to calculate the correlation function. In this work, we adopt the FFT-based approach, because we use information on galaxy clustering at large scales, $40\hMpc<r$, in our analysis (for details, see Section~\ref{Sec:FittingPrescription}).

\begin{figure}
	\includegraphics[width=\columnwidth]{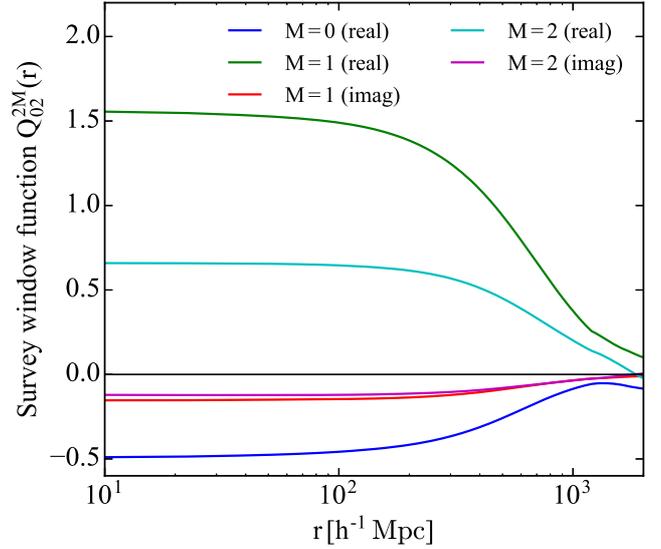}
	\caption{
BipoSH coefficients of survey window functions $Q_{20}^{2M}$ as given in equation~(\ref{Eq:Q_LM}) for CMASS NGC,
which are used to compute the masked power spectrum and correlation function given by equations~(\ref{Eq:template_pk}) and (\ref{Eq:template_xi}), respectively.
They represent survey geometry asymmetries and result in statistical anisotropic signals on the BipoSH coefficients of the power spectrum and correlation function.
}
	\label{fig:window}
\end{figure}

\begin{figure*}
	\includegraphics[width=1.0\textwidth]{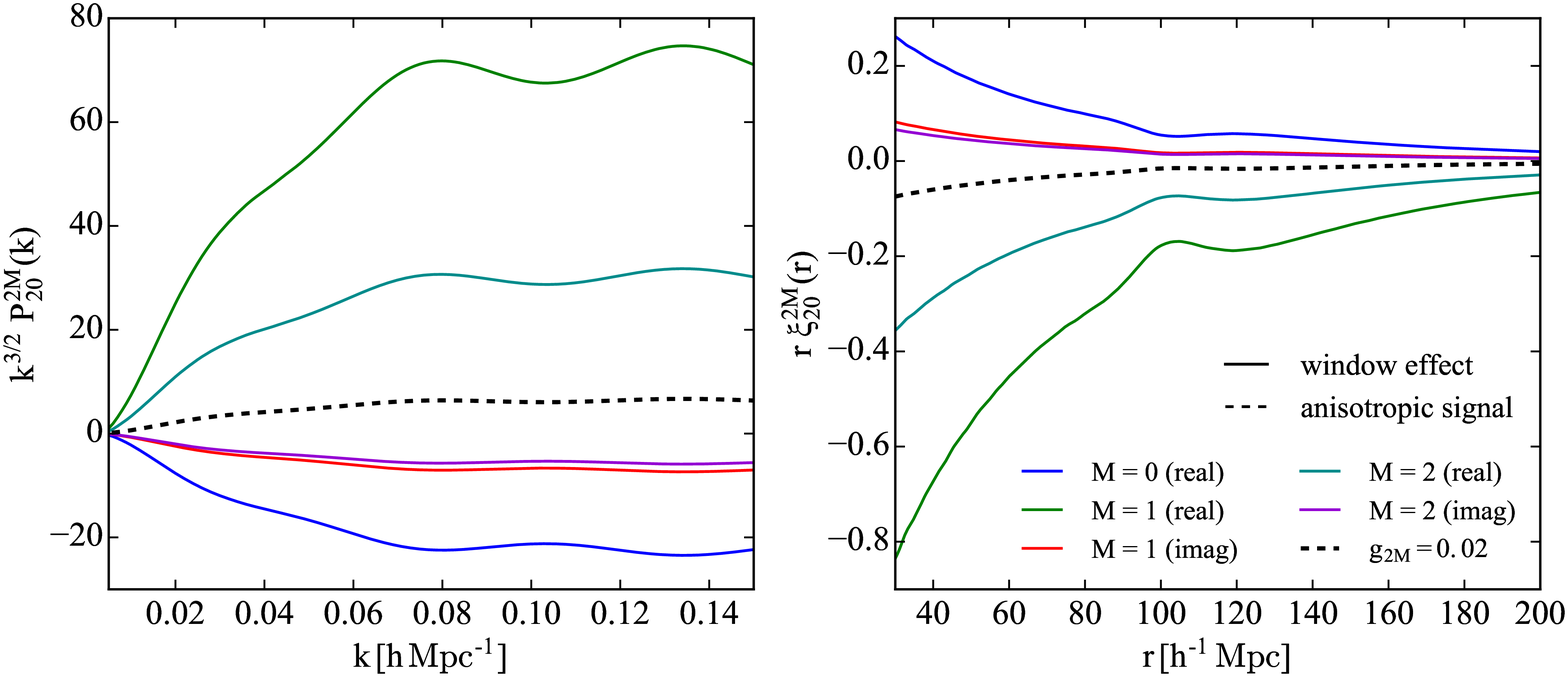}
	\caption{
BipoSH coefficients of the power spectrum $P_{20}^{2M}$ (left) and the correlation function $\xi_{20}^{2M}$ (right) for CMASS NGC as given in equations~(\ref{Eq:template_pk}) and (\ref{Eq:template_xi}), respectively. The solid colored lines represent the window function corrections, while the dashed black lines denote the primordial anisotropic signal. This figure shows that the mimic anisotropic signals induced by the survey geometry make a significant contribution to the observed BipoSH coefficients.
	}
	\label{fig:pkxi_window}
\end{figure*}

The correlation function is not usually normalized by the factor $A$ but the number of pairs of a random distribution at each bin $r$ ~\citep{Landy1993ApJ...412...64L}. Therefore, we finally present the following estimators of the spherical harmonic and BipoSH coefficients of the correlation function 
\begin{eqnarray}
	\widehat{\xi}_{\ell m; \ell' m'}(r) &=& \frac{\widehat{\xi}_{\ell m; \ell' m'}(r)\big|_{\rm A} }{RR(r)}, \nonumber \\
	\widehat{\xi}_{\ell \ell' }^{\, LM}(r) &=& \frac{\widehat{\xi}_{\ell \ell' }^{\, LM}(r)\big|_{\rm A} }{RR(r)}, 
	  \label{Eq:estimator_XillLM_RR}
\end{eqnarray}
where $RR(r)$ is the auto-correlation function measured from the random catalog
\begin{eqnarray}
	RR(r) \hspace{-0.25cm} &=& \hspace{-0.25cm}\frac{1}{A} \int \frac{d^2\hat{r}}{4\pi} \int d^3x_1\int d^3x_2  
	\delta_{\rm D}\left( \vec{r} - \vec{x}_{12}\right) \bar{n}(\vec{x}_1)\, \bar{n}(\vec{x}_2) \nonumber \\
	&=& \hspace{-0.25cm}\frac{\alpha^2}{4\pi r^2 \Delta r A} \sum_{i,j}^{N_{\rm ran}}\, \delta_{\rm K}\left( r - |\vec{x}_{ij}| \right)\, w(\vec{x}_i)\,w(\vec{x}_j).
	\label{Eq:Q}
\end{eqnarray}
The above estimators, $\widehat{\xi}_{\ell\ell'}^{\, LM}$ and $\widehat{\xi}_{\ell m;\ell' m'}$, are related to the Legendre coefficients estimator as follows
\begin{eqnarray}
	  \widehat{\xi}_{\ell}(r) = 
	  \widehat{\xi}_{\ell \ell}^{\, 00}(r) = 
	  \frac{1}{2\ell+1} \sum_{m=-\ell}^{\ell}(-1)^{m}\,\widehat{\xi}_{\ell,-m; \ell m}(r).
\end{eqnarray}

\section{SURVEY WINDOW FUNCTIONS}
\label{Sec:SurveyWindowFunctions}

In this section, we discuss the effects of survey geometry asymmetries on our statistics, the BipoSH coefficients. A treatment of the survey geometry effects for the Legendre coefficients was recently developed by~\citet{Wilson2015arXiv151107799W}, based on configuration-space calculations. We extend their treatment to the BipoSH coefficients. In this section we present the main equations, leaving their full derivations in Appendix~\ref{Ap:Derivations}.

The theoretical expression of the observed density fluctuation $\delta n$ (equation~\ref{Eq:fluctuation}) is described as
\begin{eqnarray}
	\delta n(\vec{x}\,) = \bar{n}(\vec{x}\,)\left( \delta(\vec{x}\,) - \bar{\delta}\,  \right),
	\label{Eq:delta_n}
\end{eqnarray}
where $\delta(\vec{x}\,)$ is the theoretically-predicted density perturbation, and the mean density perturbation $\bar{\delta}$ is given by
\begin{eqnarray}
	\bar{\delta} = \frac{1}{N_{\rm wg}} \int d^3x\, \bar{n}(\vec{x}\,)\, \delta(\vec{x}\,)
	\label{Eq:mean_density_perturbation}
\end{eqnarray}
with $N_{\rm wg}$ being the weighted total number of galaxies $N_{\rm wg} = \int d^3x\, \bar{n}(\vec{x}) = \sum_i^{N_{\rm gal}}w(\vec{x}_i)$. As the mean number density $\bar{n}(\vec{x}\,)$ is estimated from a random distribution in a finite survey volume, which is given by equation~(\ref{Eq:density_ran}), it behaves as the survey mask. The mean density perturbation $\bar{\delta}$, the so-called integral constraint~\citep{Peacock1991MNRAS.253..307P}, comes from the difference between the measured mean density from a finite survey volume and the true value. Equation~(\ref{Eq:delta_n}) satisfies $\int d^3x\,\delta n(\vec{x}\,)= \delta n(\vec{k}=0) = 0$ due to the integral constraint term. Using equation~(\ref{Eq:delta_n}), we obtain
\begin{eqnarray}
	  \left\langle \delta n(\vec{x}_1)\,\delta n(\vec{x}_2) \right\rangle
	  = \bar{n}(\vec{x}_1) \bar{n}(\vec{x}_2) \left[ \xi(\vec{x}_{12}, \hat{x}_{1})- \left\langle \bar{\delta}^2\right\rangle  \right],
	  \label{Eq:xi_obs_theory}
\end{eqnarray}
where we used an approximation $\langle \delta(\vec{x}) \bar{\delta}\rangle \approx \langle \bar{\delta}^2\rangle$\footnote{The treatment which does not rely on this approximation may be useful to improve the accuracy, but we leave it for future work.}. In the above expression, we do not use the global LOS direction $\hat{n}$ used in Section~\ref{Sec:Theory} but a local LOS direction, $\hat{x}_1\approx\hat{x}_2$, under the local plane parallel approximation. For the correlation function, the BipoSH expansion derived in Section~\ref{Sec:Theory} holds true even if we replace $\hat{n}$ by $\hat{x}_{1}$.
\footnote{The discussion in this section will hold true even if we do not use a local LOS direction $\hat{x}_{1}$ but the other definition of the LOS direction, e.g. the unit vector $\hat{n}_{12}$ of $\vec{n}_{12}=(\vec{x}_1+\vec{x}_2)/2$.}

Since we observe the BipoSH coefficients estimators, $\widehat{P}_{\ell \ell'}^{\, 2M}(k)$ and $\widehat{\xi}_{\ell \ell'}^{\, 2M}(r)$, given in Section~\ref{Sec:Methodology}, we should compute the ensemble averages of them, $\big\langle \widehat{P}_{\ell \ell'}^{\, 2M}(k) \big\rangle$ and $\big\langle \widehat{\xi}_{\ell \ell'}^{\, 2M}(r) \big\rangle$, to construct theoretical models for the BipoSH coefficients including the survey geometry effect. For that purpose, we first compute the ensemble average of $\widehat{\xi}_{\ell \ell' }^{\, LM}(r)|_{\rm A}$ (equation~\ref{Eq:estimator_XillLM_A}):
\begin{eqnarray}
	\hspace{-0.25cm}
	\left\langle\widehat{\xi}_{\ell \ell' }^{\, LM}(r)\big|_{\rm A} \right\rangle
	\hspace{-0.25cm}
	&=&  
	\hspace{-0.25cm}
	\frac{(2L+1)(2\ell+1)(2\ell'+1)H_{\ell\ell'}^L}{A} \int \frac{d^2\hat{r}}{4\pi} \nonumber \\
	&\times&
	\hspace{-0.25cm}
	\int d^3x_1\int d^3x_2  \delta_{\rm D}\left( \vec{r} - \vec{x}_{12}\right)\,  S_{\ell\ell'}^{LM*}(\hat{r},\hat{x}_{1}) \nonumber \\
	&\times&
	\hspace{-0.25cm}
	\bar{n}(\vec{x}_1)\, \bar{n}(\vec{x}_2)
		  \left[ \xi(\vec{r},\hat{x}_{1}) - \left\langle \bar{\delta}^2 \right\rangle \right],
	\label{Eq:xiA_average}
\end{eqnarray}
where $\xi(\vec{r},\hat{x}_{1})$ can be expanded in BipoSHs (equation~\ref{Eq:XiLMll_from_Pllmm}). Second, we define the BipoSH coefficients of survey window functions as
\begin{eqnarray}
	Q_{\ell\ell'}^{\,LM}(r) &\equiv& (2L+1)\, H_{\ell\ell'}^L\, (-1)^{M} \nonumber \\
	&\times& \sum_{mm'} \left( \begin{smallmatrix} \ell & \ell' & L \\ m & m' & -M \end{smallmatrix}  \right)\, Q_{\ell m;\ell'm'}(r),
	\label{Eq:Q_LM}
\end{eqnarray}
where the spherical harmonic coefficients of the window function are given by
\begin{eqnarray}
	Q_{\ell m;\ell' m'}(r) 
	\hspace{-0.25cm}&=& \hspace{-0.25cm}
	\frac{(2\ell+1)(2\ell'+1)}{A} \int \frac{d^2\hat{r}}{4\pi}\, y_{\ell m}^{*}(\hat{r}) 
	\int d^3x_1\int d^3x_2  \nonumber \\
	&\times&\hspace{-0.25cm}
	\delta_{\rm D}\left( \vec{r} - \vec{x}_{12}\right)\,  y_{\ell' m'}^{*}(\hat{x}_{1})\, \bar{n}(\vec{x}_1)\, \bar{n}(\vec{x}_2).
\end{eqnarray}
These spherical harmonic and BipoSH coefficients are related to the Legendre coefficients as follows
\begin{eqnarray}
	Q_{\ell}(r) = Q^{00}_{\ell\ell}(r) = \frac{1}{2\ell+1}\sum_{m=-\ell}^{\ell}\, (-1)^{m}\, Q_{\ell,-m; \ell m}(r),
	\label{Eq:Q_ELL}
\end{eqnarray}
where the monopole $Q_{0}(r)$ is equivalent to the function $RR(r)$ given by equation~(\ref{Eq:Q}). Third, we compute the integral constraint term $\langle \bar{\delta}^2 \rangle$
(Appendix~\ref{Ap:DerivationOfEquation1})
\begin{eqnarray}
	\langle \bar{\delta}^2 \rangle &=& \frac{4\pi}{V}\int dr r^2 \left[\sum_{\ell}\frac{1}{2\ell+1} Q_{\ell}(r) \xi_{\ell}(r)\right],
	\label{Eq:Integral_constraint}
\end{eqnarray}
where the survey volume $V$ is estimated as $N_{\rm wg}^2/A$. Fourth, by substituting equations~(\ref{Eq:XiLMll_from_Pllmm}) and (\ref{Eq:Q_LM}) into equation~(\ref{Eq:xiA_average}), we derive a linear combination of the correlation function and the window function, especially for $\big\langle \widehat{\xi}_{20}^{\, 2M}|_{\rm A} \big\rangle$ 
(Appendix~\ref{Ap:DerivationOfEquation2})
\begin{eqnarray}
	 \left\langle \widehat{\xi}_{20}^{\ 2M}(r)\big|_{\rm A}  \right\rangle &=&
	 Q_0(r)\, \xi_{20}^{2M}(r) + Q_{20}^{2M}(r)\,\left[ \xi_0(r) - \left\langle \bar{\delta}^2 \right\rangle \right] \nonumber \\
	&+& \frac{1}{5}\, \left[ Q_{02}^{2M}(r) + Q_{22}^{2M}(r) + Q_{42}^{2M}(r)  \right]\, \xi_{2}(r) \nonumber \\
	&+& \frac{1}{9}\, \left[  Q_{24}^{2M}(r) + Q_{44}^{2M}(r) \right]\, \xi_{4}(r) \nonumber \\
	&+& \cdots,
	\label{Eq:full_xi}
\end{eqnarray}
where the first term corresponds to the signal of the statistical anisotropy, and the other terms arise from the survey geometry anisotropy. While we only use linear theory in our analysis, non-linear theories will yield additional higher Legendre coefficients other than the above expression, namely $\xi_{\ell\geq 6}$. Fifth, by keeping the dominant terms in equation (\ref{Eq:full_xi}), we have
\begin{eqnarray}
	\left\langle \widehat{\xi}_{20}^{\ 2M}(r)\big|_{\rm A}  \right\rangle	
	\approx Q_0(r)\, \xi^{2M}_{20}(r) + \frac{1}{5}\, Q_{02}^{2M}(r)\, \xi_2(r).
	\label{Eq:ap_xi}
\end{eqnarray}
We have checked and confirmed that the relative difference between equations~(\ref{Eq:full_xi}) and~(\ref{Eq:ap_xi}) is within $10\%$ on the scales of interest, $40\hMpc<r<200\hMpc$. Since the errors in the observed BipoSH coefficients estimated from the BOSS data in Section~\ref{Sec:CovarianceMatrices} are significantly larger than this relative difference, this approximation will yield negligibly small changes in the final results. Finally, we derive the masked power spectrum from equation~(\ref{Eq:XI_TO_PK}),
\begin{eqnarray}
	  \left\langle \widehat{P}_{20}^{\, 2M}(k) \right\rangle \hspace{-0.25cm}
	  &=&
	\hspace{-0.25cm}
	  - 4\pi\, \int dr\, r^2\, j_2(kr) \nonumber \\
	&\times& \left[ Q_0(r)\, \xi_{20}^{2M}(r) + \frac{1}{5}\, Q_{02}^{2M}(r)\, \xi_{2}(r) \right]
	\label{Eq:template_pk}
\end{eqnarray}
and the masked correlation function from equation~(\ref{Eq:estimator_XillLM_RR}),
\begin{eqnarray}
	 \left\langle \widehat{\xi}_{20}^{\ 2M}(r) \right\rangle = 
	  \xi^{2M}_{20}(r) + \frac{1}{5}\,\frac{Q_{02}^{2M}(r)}{Q_0(r)}\, \xi_2(r).
	  \label{Eq:template_xi}
\end{eqnarray}
We use these equations~(\ref{Eq:template_pk}) and (\ref{Eq:template_xi}) as a template model to fit the measurements of $\widehat{P}^{\, 2M}_{20}$ and $\widehat{\xi}^{\, 2M}_{20}$ in Section \ref{Sec:Results}. The theoretical predictions of $\xi_{20}^{2M}$ and $\xi_{\ell=0,2}$ are given by equations~(\ref{Eq:model_xi}) and (\ref{Eq:xi_Legendre}), where the linear matter power spectrum used in this paper is generated with \textit{CLASS}~\citep{Lesgourgues:2011re}.

Figure~\ref{fig:window} shows the BipoSH coefficients of the survey window functions $Q_{02}^{2M}$ for all $M$ used in our analysis for CMASS NGC. They should be zero at larger scales than the survey volume ($\sim 1\hGpc$), while on small scales where the survey edge effects no longer matter, they become constant. These properties are similar to the monopole of the window function $Q_0$ (see e.g. Figure $2$ in~\cite{Beutler2016arXiv160703150B}). 

Figure~\ref{fig:pkxi_window} shows the BipoSH coefficients of the power spectrum $P_{20}^{2M}$ (left panel) and correlation function $\xi_{20}^{2M}$ (right panel) for CMASS NGC. The solid colored lines represent the mimic anisotropic signals caused by the survey geometry, which are given by the second terms of equations~(\ref{Eq:template_pk}) and (\ref{Eq:template_xi}), while the dashed black lines denote the primordial anisotropic signal described by the first terms, where we adopt $g_{2M}=0.02$ as a typical value. We find that the effects of survey geometry on the BipoSH coefficients appear even on small scales and make a significant contribution to the observed power spectrum and correlation function.

\section{MEASUREMENTS}
\label{Sec:Measurements}

\begin{figure*}
	\includegraphics[width=1.0\textwidth]{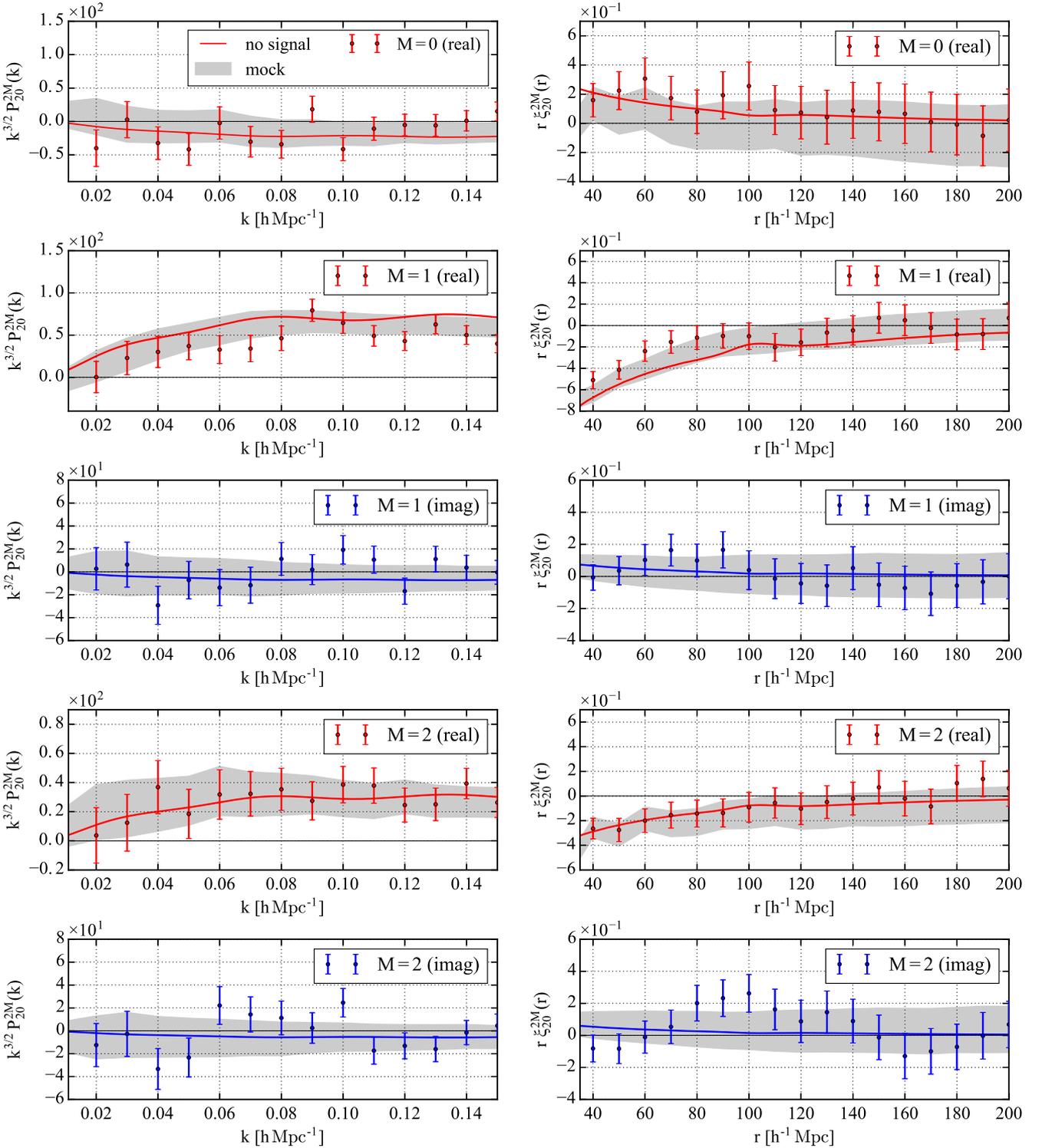}
	\caption{
	BipoSH coefficients of the power spectrum $P_{20}^{2M}$ (left panels) and the correlation function $\xi_{20}^{2M}$ (right panels) measured from CMASS NGC. The errorbars on the data points are derived from the $1\sigma$ errors measured from the QPM mocks. The gray shaded regions are the measurements from the QPM mocks, which do not include statistically anisotropic features, with the $1\sigma$ errors. The solid lines denote the predictions of the linear theory assuming no anisotropic signal, namely $g_{2M}=0$, which are the same as the solid colored lines in Figure~\ref{fig:pkxi_window}. The predictions from the QPM mocks and the theory only include the effects of survey geometry asymmetries and are in excellent agreement with the observations. Therefore, these figures show that the observed BipoSH coefficients, $P_{20}^{2M}$ and $\xi_{20}^{2M}$ for all $M$, can be sufficiently explained by the survey geometry effects.
	}
	\label{fig:pkxi_CMASS_North}
\end{figure*}

\begin{figure*}
	\includegraphics[width=1.0\textwidth]{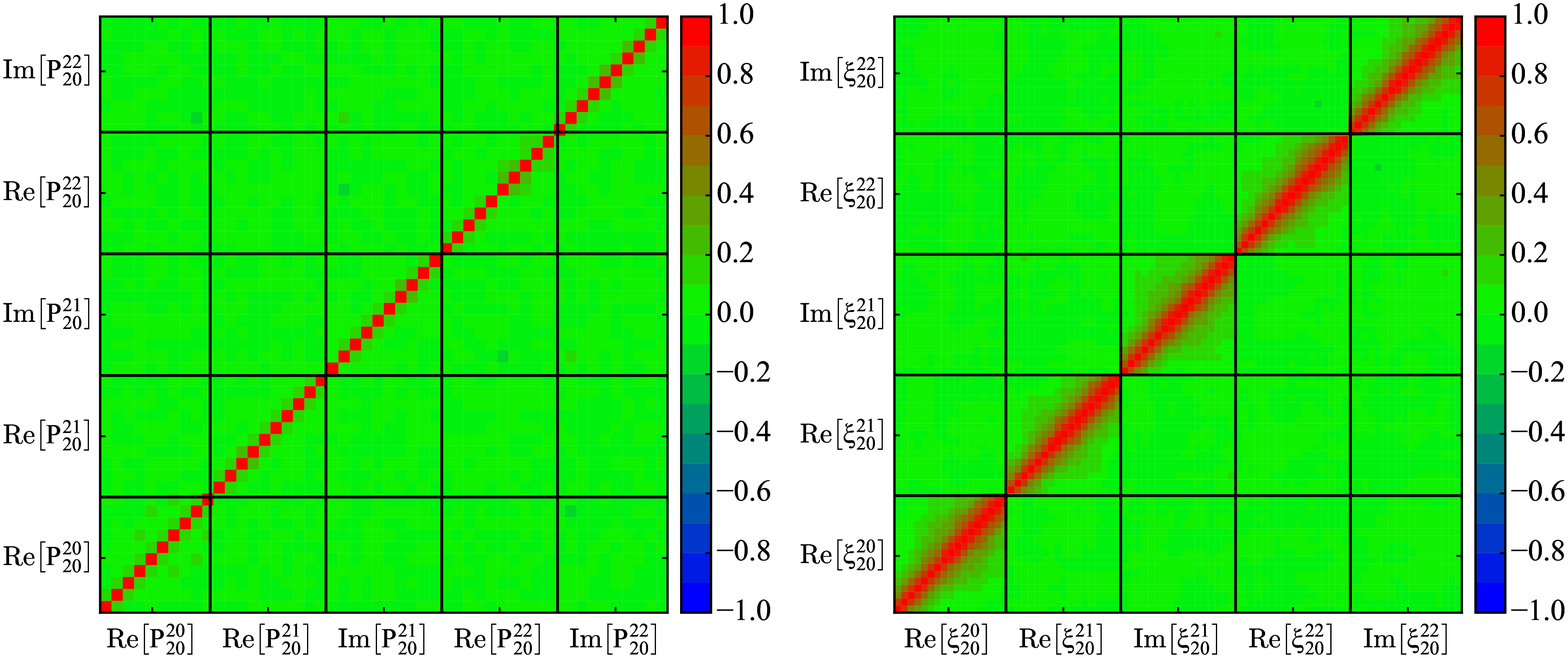}
	\caption{
Correlation coefficient matrices of both the power spectrum (left) and the correlation function (right), estimated from the QPM mocks for CMASS NGC. In the left panel, each block separated by black lines includes $k$-bins between $k=0.01\, \mathchar`-\, 0.1\hk$, while in the right panel each block includes $r$-bins between $r=40\, \mathchar`-\, 200\hMpc$. The color indicates the level of correlation, where red represents high correlation, and green denotes low correlation. These figures indicate a very weak correlation between different two $M$ modes of the BipoSH coefficients in both the Fourier- and configuration-space analyses.}
	\label{fig:covariance}
\end{figure*}

\subsection{Power spectrum and correlation function}

In our analysis, we compute the power spectrum and the correlation function using the Fast Fourier Transform in the West (FFTW)\footnote{ \url{http://fftw.org}}. We define the Cartesian coordinates $\vec{x}=\left( x,y,z \right)$ with $z$ being the axis toward the north pole and place the LOWZ and CMASS samples in a cuboid of dimensions $\left( L_x,\, L_y,\, L_z \right)\, [\hMpc]$, where $\left( L_x,\, L_y,\, L_z \right)$ $ = \left( 2400,\, 4200,\, 2400 \right)$ for CMASS NGC, $\left( 2600,\, 3400,\, 2000 \right)$ for CMASS SGC, $\left( 2300,\, 4000,\, 2300 \right)$ for LOWZ NGC, and $\left( 2400,\, 3200,\, 1700 \right)$ for LOWZ SGC. We then distribute the CMASS and LOWZ galaxies on the FFT grid using the TSC assignment function with a $512$ grid on an axis. This corresponds to a grid-cell resolution of $\sim 5\hMpc$ for both CMASS and LOWZ. 

To estimate numerical convergence errors in FFT computations, we measure the power spectrum and the correlation function for CMASS NGC with two FFT grids, $512$ and $1024$, on an axis. We then define a fractional quantity, the difference between the two power spectra/correlation functions divided by the standard deviation of the power spectrum/correlation function estimated with the $512$ grid in Section~\ref{Sec:CovarianceMatrices}. We have checked that on the scales of interests, $0.01\hk < k < 0.10\hk$ and $40\hMpc < r < 200\hMpc$, the fractional differences for the power spectrum and correlation function are within $2\%$ and $20\%$, respectively. Since the correlation function that is computed by the FFT-based approach will cause larger systematic biases than those induced by the power spectrum, we adopt the constraint on $g_{2M}$ derived from the Fourier-space analysis as the main results.\footnote{Although we use the power spectrum as the main analysis, the fractional difference of $20\%$ for the correlation function will not significantly affect the constraint on the quadrupolar parameter, $g_{2M}$, due to larger error on $g_{2M}$ than its signal, i.e. no evidence for violation of SI.}

Figure~\ref{fig:pkxi_CMASS_North} presents the measurements of both $P_{20}^{2M}$ (left panels) and $\xi_{20}^{2M}$ (right panels) for CMASS NGC using the estimators detailed in Sections~\ref{Sec:PowerSpectrum} and~\ref{Sec:TwoPointCorrelationFunction}, where their real and imaginary parts are shown by red and blue symbols, respectively.  The solid lines are the fiducial models under the no anisotropic signal hypothesis $g_{2M}=0$, which only include survey geometry corrections discussed in Section~\ref{Sec:SurveyWindowFunctions}. 

\begin{figure*}
	\includegraphics[width=1.0\textwidth]{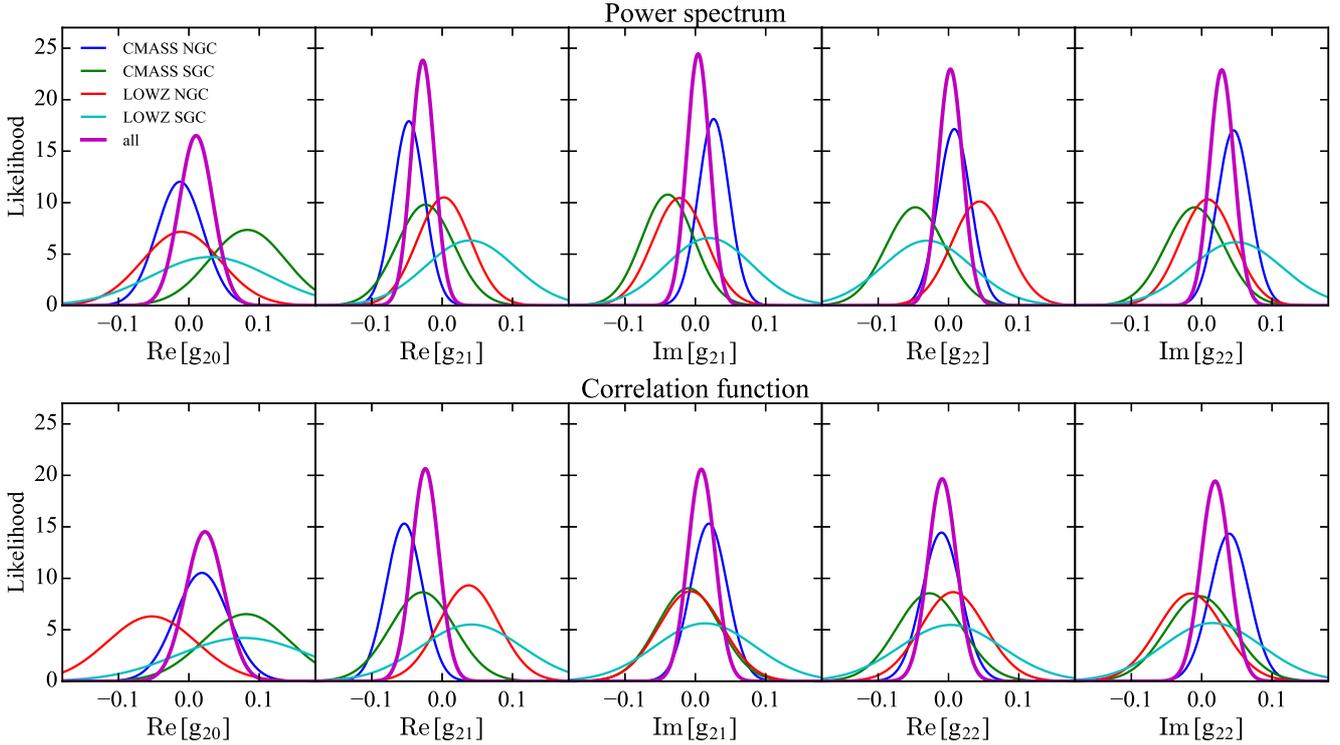}
	\caption{Likelihood functions for $g_{2M}$ in Fourier (top panels) space and configuration (bottom panels) space.
	The colored lines in each panel show the likelihoods for four galaxy samples, CMASS NGC (blue), CMASS SGC (green), LOWZ NGC (red), and LOWZ SGC (cyan),
	and the likelihood computed by combining the four samples (purple).
	We treat the parameters $g_{2M}$ for all $M$ as statistically independent quantities and estimate the likelihoods for the four samples separately.
	}
	\label{fig:likelihood_g2M}
\end{figure*}

\subsection{Covariance matrix}
\label{Sec:CovarianceMatrices}

Once the power spectrum and correlation function are observed, it is necessary to estimate the error, namely the covariance matrix. One of the best ways to derive the covariance matrix is to utilize a number of mock catalogs made for a given sample. As shown in S17, the monopole of the power spectrum, $P_0$, has a dominant contribution to the covariance matrix of the BipoSH coefficient $P_{20}^{2M}$. We thus expect that mock catalogs which do not include statistically anisotropic features are suited to use for the covariance estimate. 

In this paper, we use the $999$ Quick-Particle-Mesh (QPM) mock catalogs~\citep{White2014MNRAS.437.2594W} that are based on low-resolution particle mesh simulations, in combination with the Halo Occupation Distribution (HOD) technique to populate the resolved halos with galaxies (see e.g. \cite{Tinker2012ApJ...745...16T}). The QPM scheme incorporates observational effects including the survey selection window and fiber collisions. For the QPM mocks, the simulation outputs are at $z=0.55$ for CMASS and $z=0.40$ for LOWZ. The fiducial cosmology for these mocks assumes a $\Lambda$CDM cosmology with $(\Omega_{\Lambda}, \Omega_m,\Omega_b,\sigma_8,h,n_s)=(0.71,0.29,0.0458,0.80,0.7,0.97)$.

Using the QPM mocks, the covariance matrix for the statistic of interest $y$ (either the power spectrum $P$ or correlation function $\xi$) is given by
\begin{eqnarray}
	C_{ij} = \frac{1}{N_{\rm mock}-1} \sum_{n}^{N_{\rm mock}} \left( y_i^{n} - \bar{y}_i \right) \left( y_j^n - \bar{y}_j \right),
	\label{Eq:Cij}
\end{eqnarray}
where $N_{\rm mock}=999$, $y_i^n$ is the $i$th binned value of the statistic $y$ obtained from the $n$th mock, and $\bar{y}_i$ is the mean value over the mocks, given by $\bar{y}_i = (1/N_{\rm mock})\sum_n^{N_{\rm mock}}y_i^n$. We denote the power spectrum vector and the correlation function vector as $\vec{P} = \{ {\rm Re}\left[ P_{20}^{20} \right], {\rm Re}\left[ P_{20}^{21} \right], {\rm Im}\left[ P_{20}^{21} \right], {\rm Re}\left[ P_{20}^{22} \right], {\rm Im}\left[ P_{20}^{22} \right]\}$ and $\vec{\xi} = \{ {\rm Re}\left[ \xi_{20}^{20} \right], {\rm Re}\left[ \xi_{20}^{21} \right], {\rm Im}\left[ \xi_{20}^{21} \right], {\rm Re}\left[ \xi_{20}^{22} \right], {\rm Im}\left[ \xi_{20}^{22} \right]   \}$, where each component of $\vec{P}$ ($\vec{\xi}$) includes $k$-bins ($r$-bins). We then estimate the covariance matrices for the power spectrum and correlation function by replacing the vector $\vec{y}$ in equation~(\ref{Eq:Cij}) by $\vec{P}$ and $\vec{\xi}$, respectively. We only need the positive $M$ modes of $P_{20}^{2M}$ and $\xi_{20}^{2M}$ due to the reality condition, $P_{20}^{2M*} = (-1)^{M}P_{20}^{2, -M}$ and $\xi_{20}^{2M*} = (-1)^{M}\xi_{20}^{2, -M}$.

The correlation coefficient matrix $r_{ij}$ is defined as $r_{ij} = C_{ij}/(C_{ii}C_{jj})^{1/2}$. Figure~\ref{fig:covariance} displays the correlation matrices of the power spectrum (left) and correlation function (right) for CMASS NGC. Each panel shows a matrix with five horizontal and vertical division lines that divide the matrix into $5\times5$ blocks. Each block in the left panel includes $k$-bins between $k = 0.01\, \mathchar`-\, 0.1\hk$ with the bin width $\Delta k = 0.01\hk$, and each block in the right panel includes $r$-bins between $r = 40\, \mathchar`-\, 200\hMpc$ with $\Delta r=10\hMpc$. We observe that almost all elements in each off-diagonal block are less than $0.1$ for the four samples, CMASS NGC, CMASS SGC, LOWZ NGC, and LOWZ SGC, in both the Fourier- and configuration-space analyses. This fact indicates that there are  negligibly small correlations between different two $M$ modes of the BipoSH coefficients on the scales of interest in our analysis. 

The error bars shown in Figure \ref{fig:pkxi_CMASS_North} are the standard deviation obtained by the square root of the diagonal components, $C_{ii}^{1/2}$. The gray shaded regions in Figure \ref{fig:pkxi_CMASS_North} are the measurements of $P_{20}^{2M}$ and $\xi_{20}^{2M}$ from the QPM mocks with the $1\sigma$ errors, which do not include the primordial anisotropic signal.

S17 has shown that there is no correlation between different $M$ modes of the BipoSH coefficients in linear theory. This characteristic feature is consistent with the covariance estimate from the QPM mocks. In Appendix~\ref{Ap:StandardDeviation}, we compare the standard deviation (the square root of the diagonal elements of the covariance matrix) of $P_{20}^{2M}$ computed by the linear theory developed in S17 with that estimated from the QPM mocks. We find an excellent agreement between the results from the theory and the mock, which validates both of the Fisher matrix computations performed by S17 and the error estimates in this paper.

As the estimated covariance matrix $C_{ij}$ in Section~\ref{Sec:CovarianceMatrices} is inferred from a set of mocks, its inverse $C^{-1}_{ij}$ is biased due to the limited number of realizations. We account for this effect by rescaling the inverse covariance matrix by the factor of equation~(17) of~\citet{Hartlap2007A&A...464..399H}. We measure the standard $\chi^2$ with this rescaled inverse covariance matrix. In addition to the Hartlap factor, we propagate the error in the covariance matrix to the error on $g_{2M}$ by scaling the variance for $g_{2M}$ by the factor of equation~(18) of~\cite{Percival2014MNRAS.439.2531P}.

\section{ANALYSIS}
\label{Sec:Results}

\begin{table*}
\centering
\begin{tabular}{crrrrr}
\hline\hline
\multicolumn{6}{c}{Power spectrum} \\
\hline
$g_{2M}/10^{-2}$ & CMASS NGC & CMASS SGC & LOWZ NGC & LOWZ SGC & \multicolumn{1}{c}{\hspace{0.3cm}All} \\
\hline  
${\rm Re}\,[g_{20}]$ & $-0.81\pm3.33$ & $ 8.81\pm5.44$ & $-0.40\pm5.58$ & $ 4.00\pm8.48 $ & $ 1.57\pm2.42$\\
${\rm Re}\,[g_{21}]$ & $-4.78\pm2.24$ & $-2.29\pm4.07$ & $ 0.03\pm3.82$ & $ 4.19\pm6.33 $ & $-2.79\pm1.68$\\
${\rm Im}\,[g_{21}]$ & $ 2.61\pm2.21$ & $-3.95\pm3.71$ & $-2.24\pm3.84$ & $ 1.98\pm6.10 $ & $ 0.40\pm1.64$ \\
${\rm Re}\,[g_{22}]$ & $ 1.16\pm2.33$ & $-4.62\pm4.19$ & $ 4.87\pm3.97$ & $-2.88\pm6.34 $ & $ 0.57\pm1.74$ \\
${\rm Im}\,[g_{22}]$ & $ 4.51\pm2.35$ & $-1.04\pm4.19$ & $ 0.78\pm3.88$ & $ 4.93\pm6.50 $ & $ 2.82\pm1.74$ \\
\hline  
\end{tabular}

\centering
\begin{tabular}{crrrrr}
\hline\hline
\multicolumn{6}{c}{Correlation function} \\
\hline
$g_{2M}/10^{-2}$ & CMASS NGC & CMASS SGC & LOWZ NGC & LOWZ SGC & \multicolumn{1}{c}{\hspace{0.3cm}All}\\
\hline
${\rm Re}\,[g_{20}]$ & $ 2.11\pm3.81$ & $ 9.06\pm6.30$ & $-4.93\pm6.37$ & $8.21\pm9.55$ & $ 2.65\pm2.78 $ \\
${\rm Re}\,[g_{21}]$ & $-5.44\pm2.62$ & $-3.00\pm4.64$ & $ 3.71\pm4.31$ & $4.24\pm7.28$ & $-2.46\pm1.94 $ \\
${\rm Im}\,[g_{21}]$ & $ 1.94\pm2.62$ & $-0.79\pm4.36$ & $-0.70\pm4.60$ & $1.38\pm7.14$ & $ 0.88\pm1.94 $ \\
${\rm Re}\,[g_{22}]$ & $-0.79\pm2.78$ & $-2.79\pm4.83$ & $ 0.94\pm4.64$ & $0.41\pm7.34$ & $-0.72\pm2.05 $ \\
${\rm Im}\,[g_{22}]$ & $ 3.91\pm2.80$ & $-0.37\pm4.67$ & $-1.62\pm4.72$ & $1.50\pm7.10$ & $ 1.84\pm2.05 $ \\
\hline  
\end{tabular}
\caption{Mean values of $g_{2M}$ with their standard deviations. The upper and lower tables are for the Fourier- and configuration-space analyses, respectively.
From second to fifth columns, the constraints on $g_{2M}$ are given for four galaxy samples, CMASS-NGC, CMASS-SGC, LOWZ-NGC, and LOWZ-SGC.
The sixth column presents the results obtained by combining all the samples, showing that the quadrupolar parameters $g_{2M}$ for all $M$ are of zero within the $2\sigma$ confidence level.
	}
\label{Table:constraints_g2M}
\end{table*}

\begin{table*}
\centering
\begin{tabular}{crrrrr}
\hline\hline
   & 
\multicolumn{1}{c}{CMASS NGC} & 
\multicolumn{1}{c}{CMASS SGC} & 
\multicolumn{1}{c}{LOWZ NGC} & 
\multicolumn{1}{c}{LOWZ SGC} & 
\multicolumn{1}{c}{All}   \\
\hline  
Power Spectrum & 
$-0.14<g_*<0.09$ &
$-0.14<g_*<0.18$ & 
$-0.12<g_*<0.11$ & 
$-0.18<g_*<0.19$ & 
$-0.093<g_*<0.079$ \\
Correlation function &
$-0.14<g_*<0.10 $ & 
$-0.13<g_*<0.18 $ & 
$-0.16<g_*<0.13 $ & 
$-0.19<g_*<0.23 $ & 
$-0.088<g_*<0.085$ \\
\hline  
\end{tabular}
\caption{Limits on $g_*$ with a $95\%$ confidence level after marginalizing over the preferred direction $\hat{p}$.
}
\label{Table:g*}
\end{table*}

\subsection{Fitting prescription}
\label{Sec:FittingPrescription}

We perform a standard likelihood analysis, where the likelihood function is computed as $L \propto \exp\left( -\chi^2/2 \right)$ and the $\chi^2$-statistics is given by $\chi^2 = \left( \textbf{d}^{T}-\textbf{m}^{T}  \right)\, \textbf{C}^{-1}\,  \left( \textbf{d}-\textbf{m}  \right)$ with the data vector $\textbf{d}$, model vector $\textbf{m}$, and covariance matrix $\textbf{C}$. 
In our analysis, the data vector is the power spectrum vector $\vec{P}$ or the correlation function vector $\vec{\xi}$, and the covariance matrix $\textbf{C}$ of each $\vec{P}$ and $\vec{\xi}$ is given in Section~\ref{Sec:CovarianceMatrices}. The model vector is computed by the ensemble average of the data vector, $\textbf{m} = \left\langle \textbf{d}\right \rangle$, as discussed in Section~\ref{Sec:SurveyWindowFunctions}. We fit the BipoSH coefficients of the power spectrum, $P_{20}^{2M}$, and correlation function, $\xi_{20}^{2M}$, with the templates respectively given by equations~(\ref{Eq:template_pk}) and (\ref{Eq:template_xi}) with the quadrupolar parameters $g_{2M}$ for all $M$ being free parameters. We fix the other parameters, the linear growth rate $f\sigma_8$, the linear bias $b\sigma_8$ and the cosmological parameters (see Section~\ref{Sec:Introduction}), using the Planck and BOSS results~\citep{Planck2016A&A...594A..13P,Gil-Marn2016MNRAS.460.4188G}.

To determine the fitting ranges of the power spectrum and correlation function, we compute the $\chi^2$ divided by the number of degrees of freedom (d.o.f), as a function of the maximum wavenumber $k_{\rm max}$ in Fourier space and the minimum comoving distance $r_{\rm min}$ in configuration space. We fix the minimum wavenumber $k_{\rm min}=0.01\hk$ and the maximum comoving distance $r_{\rm max}=200\hMpc$. We calculate the $\chi^2$ using the QPM mocks and the model with $g_{2M}=0$, clarifying the scales where the linear theory approximation breaks in our analysis. We find that $\chi^2/{\rm d.o.f.}$ starts to significantly depart from unity at $k_{\rm max}\sim0.1\hk$ and $r_{\rm min}\sim40\hMpc$ for all the four galaxy samples. Therefore, we decide to use the fitting ranges of $k=0.01\, \mathchar`-\, 0.1\hk$ and $r=40\, \mathchar`-\, 200\hk$ and bin the power spectrum and correlation function in bins of $\Delta k = 0.01\hk$ and $\Delta r = 10\hMpc$.

Since the correlation between different $M$ modes of each of $P_{20}^{2M}$ and $\xi_{20}^{2M}$ was shown to be very weak in Section~\ref{Sec:CovarianceMatrices}, we treat the $M$ modes of each $P_{20}^{2M}$ and $\xi_{20}^{2M}$ as statistically independent quantities in our analysis. This treatment implies that the quadrupolar parameters $g_{2M}$ for all $M$ are statistically independent of each other. To validate this treatment, we perform a full analysis for CMASS NGC in Fourier space. Namely, we compute the likelihood function for $g_{2M}$ using the full covariance matrix of $P_{20}^{2M}$ including the correlation between their different $M$ modes and estimate the correlation coefficient matrix of $g_{2M}$. We then find that all off-diagonal elements of the correlation matrix of $g_{2M}$ are less than $0.06$, indicating no correlation between $g_{2M}$ and $g_{2M'}$ for $M\neq M'$. We expect the similar results even for the other samples and for the configuration-space analysis.

\subsection{Parameter constraints}

Figure~\ref{fig:likelihood_g2M} shows the likelihood functions of $g_{2M}$ for four galaxy samples, CMASS NGC (blue), CMASS SGC (green), LOWZ NGC (red), and LOWZ SGC (cyan), and the likelihood computed by combining these four samples (purple). We estimate the likelihood functions for the four samples separately, i.e. treat them as statistically independent samples. The top and bottom panels are the results for the power spectrum and correlation function, respectively. Since the quadrupolar parameter $g_{2M}$ is a proportionality constant in our template model, the shape of each likelihood becomes closely similar to a Gaussian distribution. Table~\ref{Table:constraints_g2M} presents the mean values of $g_{2M}$ with their standard deviations $\sigma_{2M}$, which are computed from the corresponding likelihood functions with the flat priors, $-1 \leq {\rm Re}[g_{2M}] \leq 1$ and $-1 \leq {\rm Im}[g_{2M}] \leq 1$. For all the four samples, $g_{2M}$ is consistent with zero at the $2\sigma$ level, except for ${\rm Re}\left[ g_{21} \right]$ for CMASS NGC. Combining all the samples, $g_{2M}$ for all $M$ are within $2\sigma$ of zero. We see consistency between the results from the Fourier- and configuration-space analyses.

Also of interest is a quadrupolar directional dependence of the primordial power spectrum, which is related to $g_{2M}$ as (see e.g. equation~$172$ in~\citet{Planck2016A&A...594A..20P})
\begin{eqnarray}
	  g_{2M} = \frac{8\pi}{15}\, g_*\, Y_{2M}^*(\hat{p}),
	  \label{Eq:g2M_g*}
\end{eqnarray}
where $\hat{p}$ is a preferred direction in space, and $g_*$ is a parameter characterizing the amplitude of the anisotropy.
Using constraints on $g_{2M}$, the likelihood for $g_*$ is given by
\begin{eqnarray}
	  L(g_*) \propto \int d^2\hat{p} \exp\left( -\frac{1}{2}\sum_{M} \left(  \frac{ g_{2M}-\frac{8\pi}{15}g_*Y_{2M}^*(\hat{p})}{\sigma_{2M}}  \right)^2
	  \right),
\end{eqnarray}
where we marginalize over all possible directions of $\hat{p}$. Since we find that the shape of this likelihood is deviated from a Gaussian distribution, we compute the lower and upper limits on $g_*$ with a $95\%$ confidence level (CL) using the flat prior $-1\leq g_* \leq 1$. We summarize the results in Table~\ref{Table:g*}; combining all the four galaxy samples, our limit on $g_*$ is $-0.09<g_*<0.08$ ($95\%\,{\rm CL}$). We also find that the marginalized likelihood for $\hat{p}$ has many peaks, indicating no preferred direction in the Universe. This fact is consistent with the limit on $g_*$, i.e. no evidence of departures from SI.

\subsection{Comparison with previous works}

We end this paper by summarizing results of previous works and comparing them with our limits.
The Planck CMB temperature maps provide the most stringent constraints on the anisotropy parameters, $g_{\rm 2M}$ and $g_*$, to date:
e.g., the $1\sigma$ errors on $g_{20}$ and $g_*$ are $\Delta g_{20}=1.2\times10^{-2}$~\citep{Planck2016A&A...594A..16P}
and $\Delta g_* = 0.016$~\citep{Kim2013PhRvD..88j1301K}.
The limit on $g_*$ from SDSS DR7 photometric galaxy data is $-0.41<g_*<+0.38$ with a $95\%$ confidence level~\citep{Pullen2010JCAP...05..027P}.
Comparing with our results ($\Delta g_{20}=2.4\times10^{-2}$ in Table~\ref{Table:constraints_g2M} and $-0.09<g_*<0.08$ in Table~\ref{Table:g*}),
we conclude that our limits are about two times as weak as the limits provided by Planck,
while this work does improve upon the constraints about four times as stringent as those from the SDSS DR7 photometric galaxy data

\section{CONCLUSIONS}
\label{Sec:Conclusions}

Statistical isotropy is a key feature of the standard inflation theory and needs to be tested in various experiments. For this purpose, we apply the BipoSH decomposition technique to the galaxy power spectrum and correlation function. The BipoSH formalism allows us to parameterize departures from statistical isotropy regarding the total angular momentum $L$, and the presence of statistical anisotropy produces the $L\geq 1$ modes in the BipoSH coefficients. In this work, we focus especially on the quadrupolar-type anisotropy, which is associated with the $L=2$ mode, and constrain the well-known quadrupolar anisotropy parameters, $g_{2M}$ and $g_*$ with the BipoSH coefficients extracted from the BOSS DR12 sample.

Survey geometry asymmetries potentially cause the largest systematic difference between the observed BipoSH coefficients and the intrinsic cosmological signal that we want to know. This work presents a modeling approach for predicting the BipoSH coefficients of the galaxy power spectrum and correlation function in light of the survey geometry effects. Figure~\ref{fig:pkxi_CMASS_North} shows that the anisotropic signal due to the specific survey geometry provides a sufficient explanation of the observed BipoSH coefficients in the BOSS DR12 data, implying the statistical isotropy of the Universe.

Tables~\ref{Table:constraints_g2M} and~\ref{Table:g*} summarize our constraints on the quadrupolar parameters, $g_{2M}$ and $g_*$. Combining four galaxy samples, CMASS NGC, CMASS SGC, LOWZ NGC, and LOWZ SGC, we find $g_{2M}$ for all $M$ to be of zero within the $2\sigma$ level and $-0.09<g_*<0.08$ with a $95\%$ confidence level. The spectroscopic catalogs of BOSS thus provide an improvement by a factor of about four compared with the photometric catalogs of SDSS~\citep{Abazajian:2009wr}, $-0.4<g_*<0.38\, (95\%\,{\rm CL})$ obtained by~\citet{Pullen2010JCAP...05..027P}. These results are the best constraint on the quadrupolar parameter from galaxy survey data currently, while they are still weaker than the Planck results~\citep{Planck2016A&A...594A..16P,Planck2016A&A...594A..20P}.

While we only use the linear regions, $k\leq 0.1\hk$, in our analysis, non-linear information up to e.g. $k=0.2\hk$ could further shrink the errors on $g_{2M}$ and $g_*$ by a factor of two~\citep{Shiraishi:2017wec}. To do so, we will need the additional modeling of non-linear perturbation theories and fiber collisions on small scales.

The tools and techniques that we have discussed here may be straightforwardly applied to constraining any source of statistical anisotropy, e.g. tidal forces $\tau_{ij}$ arising from the super-sample mode beyond the survey area~\citep{Akitsu:2016leq}. The BOSS data will provide the same order of errors on $\tau_{ij}$ as that associated with the quadrupolar parameter, namely $\Delta \tau_{ij} = {\cal O}\left( 10^{-2} \right)$.

All the analysis presented in this work will be directly applicable to future spectroscopic galaxy surveys, e.g. the Subaru Prime Focus Spectrograph (PFS;~\cite{Takada2014PASJ...66R...1T}), the Dark Energy Spectroscopic Instrument (DESI;~\cite{Levi2013arXiv1308.0847L}), and Euclid~\citep{Laureijs:2011gra}, which will provide much better sensitivity to $g_{2M}$ and $g_*$. We have forecast in~\citet{Shiraishi:2017wec} that PFS and Euclid could achieve the sensitivity comparable to or even better than the Planck results~\citep{Planck2016A&A...594A..16P,Planck2016A&A...594A..20P}.

\section*{ACKNOWLEDGEMENTS}

We are grateful to Yin Li for discussion. NSS and MS acknowledge financial support from Grant-in-Aid for JSPS Fellows (Nos. 28-1890 and 27-10917). We were supported in part by the World Premier International Research Center Initiative (WPI Initiative), MEXT, Japan. Numerical computations were carried out on Cray XC30 at Center for Computational Astrophysics, National Astronomical Observatory of Japan.

Funding for SDSS-III has been provided by the Alfred P. Sloan Foundation, the Participating Institutions, the National Science Foundation, and the U.S. Department of Energy Office of Science. The SDSS-III web site is \url{ttp://www.sdss3.org/}. SDSS-III is managed by the Astrophysical Research Consortium for the Participating Institutions of the SDSS-III Collaboration including the University of Arizona, the Brazilian Participation Group, Brookhaven National Laboratory, Carnegie Mellon University, University of Florida, the French Participation Group, the German Participation Group, Harvard University, the Instituto de Astrofisica de Canarias, the Michigan State/Notre Dame/JINA Participation Group, Johns Hopkins University, Lawrence Berkeley National Laboratory, Max Planck Institute for Astrophysics, Max Planck Institute for Extraterrestrial Physics, New Mexico State University, New York University, Ohio State University, Pennsylvania State University, University of Portsmouth, Princeton University, the Spanish Participation Group, University of Tokyo, University of Utah, Vanderbilt University, University of Virginia, University of Washington, and Yale University. 

\bibliographystyle{mnras}
\bibliography{ms} 

\appendix

\section{SCALE-DEPENDENCE OF THE QUADRUPOLAR MODULATION}
\label{Ap:Nneq0}

We consider the scale dependence of the quadrupolar modulation, namely $f(k)\neq 1$ in equation~(\ref{Eq:parameterized_model}). We assume $f(k)$ as a power law, $f(k) = (k/k_0)^n$ with three values of the spectral index, $n=-2$, $-1$, and $1$. Our constraints on $g_{2M}$ and $g_*$ will then depend on the pivot scale, chosen as $k_0 = 0.05\hk$ as elsewhere (e.g., \citet{Planck2016A&A...594A..20P}). Tables~\ref{Ap:Table:constraints_g2M_index} and \ref{Ap:Table:g*_index} show the constraints on $g_{2M}$ and $g_*$, respectively. 

\begin{table*}
\centering
\begin{tabular}{ccrrrrr}
\hline\hline
\multicolumn{7}{c}{Power spectrum} \\
\hline
Spectral index & $g_{2M}/10^{-2}$ & CMASS NGC & CMASS SGC & LOWZ NGC & LOWZ SGC & \multicolumn{1}{c}{\hspace{0.4cm}All} \\
\hline  
$n=-2$ & ${\rm Re}\,[g_{20}]$ & $ 0.78\pm2.01$ & $6.02\pm4.14$ & $-2.24\pm3.29$ & $ 5.66\pm6.54$ & $ 1.11\pm1.54$ \\
       & ${\rm Re}\,[g_{21}]$ & $-2.18\pm1.39$ & $3.04\pm2.77$ & $ 0.36\pm2.30$ & $ 3.01\pm4.42$ & $-0.57\pm1.06$ \\
       & ${\rm Im}\,[g_{21}]$ & $-0.18\pm1.37$ & $0.87\pm3.09$ & $-1.59\pm2.43$ & $-2.92\pm4.39$ & $-0.50\pm1.08$ \\
       & ${\rm Re}\,[g_{22}]$ & $-0.65\pm1.40$ & $2.89\pm3.22$ & $ 1.93\pm2.46$ & $ 2.99\pm4.61$ & $ 0.50\pm1.11$ \\
       & ${\rm Im}\,[g_{22}]$ & $-1.17\pm1.41$ & $4.93\pm2.98$ & $-2.42\pm2.43$ & $-2.09\pm4.51$ & $-0.65\pm1.09$ \\
\hline  
$n=-1$ & ${\rm Re}\,[g_{20}]$ & $-0.79\pm3.96$ & $ 11.1\pm6.86$ & $-1.12\pm6.77$ & $ 9.04\pm10.7$ & $ 2.08\pm2.94$ \\
       & ${\rm Re}\,[g_{21}]$ & $-6.67\pm2.69$ & $ 0.95\pm5.06$ & $ 0.43\pm4.75$ & $ 6.04\pm7.99$ & $-3.25\pm2.05$ \\
       & ${\rm Im}\,[g_{21}]$ & $ 1.23\pm2.68$ & $-3.20\pm4.78$ & $-3.61\pm4.61$ & $-1.08\pm7.57$ & $-0.64\pm2.01$ \\
       & ${\rm Re}\,[g_{22}]$ & $ 0.45\pm2.81$ & $-1.15\pm5.40$ & $ 5.31\pm4.85$ & $ 1.02\pm7.99$ & $ 1.18\pm2.13$ \\
       & ${\rm Im}\,[g_{22}]$ & $ 1.73\pm2.85$ & $ 1.45\pm5.22$ & $-2.41\pm4.72$ & $ 2.27\pm8.04$ & $ 0.88\pm2.13$ \\
\hline  
$n=+1$ & ${\rm Re}\,[g_{20}]$ & $-0.38\pm2.05$ & $ 5.33\pm3.36$ & $-0.88\pm3.44$ & $ 1.15\pm5.18 $ & $ 0.79\pm1.49$ \\
       & ${\rm Re}\,[g_{21}]$ & $-2.22\pm1.40$ & $-2.22\pm2.49$ & $-0.27\pm2.32$ & $ 2.14\pm3.76 $ & $-1.56\pm1.06$ \\
       & ${\rm Im}\,[g_{21}]$ & $ 2.29\pm1.37$ & $-2.52\pm2.27$ & $-0.76\pm2.37$ & $ 1.87\pm3.71 $ & $ 0.75\pm1.01$ \\
       & ${\rm Re}\,[g_{22}]$ & $ 0.82\pm1.45$ & $-3.98\pm2.55$ & $ 3.07\pm2.43$ & $-2.89\pm3.83 $ & $ 0.12\pm1.07$ \\
       & ${\rm Im}\,[g_{22}]$ & $ 3.50\pm1.44$ & $-0.53\pm2.54$ & $ 1.38\pm2.37$ & $ 3.73\pm3.98 $ & $ 2.37\pm1.07$ \\
\hline  
\end{tabular}

\centering
\begin{tabular}{ccrrrrr}
\hline\hline
\multicolumn{7}{c}{Correlation function} \\
\hline
Spectral index & $g_{2M}/10^{-2}$ & CMASS NGC & CMASS SGC & LOWZ NGC & LOWZ SGC & \multicolumn{1}{c}{\hspace{0.4cm}All}\\
\hline  
$n=-2$ & ${\rm Re}\,[g_{20}]$ & $-0.43\pm2.75$ & $ 4.15\pm5.07$ & $-3.31\pm4.69$ & $ 11.1\pm7.70$ & $ 0.60\pm2.07$ \\
       & ${\rm Re}\,[g_{21}]$ & $-3.67\pm1.87$ & $ 2.82\pm3.44$ & $-0.49\pm3.41$ & $ 2.41\pm5.48$ & $-1.59\pm1.43$ \\
       & ${\rm Im}\,[g_{21}]$ & $ 0.35\pm1.90$ & $-0.33\pm3.45$ & $-2.09\pm3.29$ & $-1.91\pm5.31$ & $-0.39\pm1.43$ \\
       & ${\rm Re}\,[g_{22}]$ & $-0.68\pm1.93$ & $ 1.97\pm3.84$ & $-0.66\pm3.44$ & $ 2.43\pm5.63$ & $-0.06\pm1.49$ \\
       & ${\rm Im}\,[g_{22}]$ & $-0.77\pm1.94$ & $ 3.46\pm3.52$ & $-4.04\pm3.35$ & $-3.73\pm5.63$ & $-0.87\pm1.47$ \\
\hline  
$n=-1$ & ${\rm Re}\,[g_{20}]$ & $ 0.38\pm4.24$ & $ 9.34\pm7.18$ & $-4.15\pm7.22$ & $ 13.6\pm11.1$ & $ 2.28\pm3.13$ \\
       & ${\rm Re}\,[g_{21}]$ & $-6.71\pm2.90$ & $ 0.34\pm5.20$ & $ 2.65\pm5.04$ & $ 5.55\pm8.29$ & $-2.86\pm2.18$ \\
       & ${\rm Im}\,[g_{21}]$ & $ 1.10\pm2.90$ & $-1.39\pm4.96$ & $-2.80\pm5.03$ & $-0.56\pm7.97$ & $-0.21\pm2.16$ \\
       & ${\rm Re}\,[g_{22}]$ & $-0.75\pm3.05$ & $-0.22\pm5.52$ & $ 0.42\pm5.26$ & $ 1.84\pm8.42$ & $-0.25\pm2.29$ \\
       & ${\rm Im}\,[g_{22}]$ & $ 1.75\pm3.10$ & $ 1.49\pm5.30$ & $-3.58\pm5.20$ & $-0.49\pm8.24$ & $ 0.50\pm2.28$ \\
\hline  
$n=+1$ & ${\rm Re}\,[g_{20}]$ & $ 2.86\pm2.86$ & $ 7.17\pm4.76$ & $-4.36\pm4.76$ & $4.41\pm7.06$ & $ 2.44\pm2.08$ \\
       & ${\rm Re}\,[g_{21}]$ & $-3.38\pm1.99$ & $-3.41\pm3.50$ & $ 3.28\pm3.19$ & $2.50\pm5.33$ & $-1.55\pm1.46$ \\
       & ${\rm Im}\,[g_{21}]$ & $ 1.91\pm2.02$ & $ 0.08\pm3.29$ & $ 0.95\pm3.47$ & $1.44\pm5.32$ & $ 1.33\pm1.48$ \\
       & ${\rm Re}\,[g_{22}]$ & $-0.91\pm2.11$ & $-3.20\pm3.63$ & $ 1.02\pm3.44$ & $0.41\pm5.39$ & $-0.83\pm1.55$ \\
       & ${\rm Im}\,[g_{22}]$ & $ 3.34\pm2.10$ & $-0.28\pm3.48$ & $-0.44\pm3.54$ & $1.63\pm5.18$ & $ 1.78\pm1.53$ \\
\hline  
\end{tabular}
\caption{Same as Table~\ref{Table:constraints_g2M} except for the scale-dependence of the quadrupolar modulation, $f(k) = (k/k_0)^n$ with $n=-2$, $-1$, and $1$,
where the pivot scale is chosen as $k_0=0.05\hk$.
	}
\label{Ap:Table:constraints_g2M_index}
\end{table*}

\begin{table*}

\centering
\begin{tabular}{crrrrr}
\hline\hline
\multicolumn{6}{c}{Power spectrum}   \\
\hline
Spectral index  & 
\multicolumn{1}{c}{CMASS NGC} & 
\multicolumn{1}{c}{CMASS SGC} & 
\multicolumn{1}{c}{LOWZ NGC} & 
\multicolumn{1}{c}{LOWZ SGC} & 
\multicolumn{1}{c}{All}   \\
\hline  
$n\,=\,-\,2$ & $-0.05<g_*<0.05$ & $-0.52<g_*<0.14$ & $-0.08<g_*<0.08$ & $-0.13<g_*<0.16$ & $-0.040<g_*<0.044$ \\
$n\,=\,-\,1$ & $-0.13<g_*<0.11$ & $-0.14<g_*<0.21$ & $-0.15<g_*<0.14$ & $-0.21<g_*<0.25$ & $-0.084<g_*<0.096$ \\
$n\,=\,+\,1$ & $-0.09<g_*<0.05$ & $-0.66<g_*<0.10$ & $-0.07<g_*<0.07$ & $-0.11<g_*<0.12$ & $-0.068<g_*<0.047$ \\
\hline  
\end{tabular}

\centering
\begin{tabular}{crrrrr}
\hline\hline
\multicolumn{6}{c}{Correlation function}\\   
\hline
Spectral index  & 
\multicolumn{1}{c}{CMASS NGC} & 
\multicolumn{1}{c}{CMASS SGC} & 
\multicolumn{1}{c}{LOWZ NGC} & 
\multicolumn{1}{c}{LOWZ SGC} & 
\multicolumn{1}{c}{All}   \\
\hline  
$n\,=\,-\,2$ & $-0.08<g_*<0.07 $ & $-0.11<g_*<0.12$ & $-0.11<g_*<0.11$ & $-0.16<g_*<0.22$ & $-0.056<g_*<0.055$ \\
$n\,=\,-\,1$ & $-0.14<g_*<0.11 $ & $-0.14<g_*<0.18$ & $-0.16<g_*<0.15$ & $-0.21<g_*<0.30$ & $-0.086<g_*<0.093$ \\
$n\,=\,+\,1$ & $-0.10<g_*<0.62 $ & $-0.11<g_*<0.14$ & $-0.12<g_*<0.10$ & $-0.14<g_*<0.16$ & $-0.072<g_*<0.066$ \\
\hline  
\end{tabular}

\caption{
Same as Table~\ref{Table:g*} except for the scale-dependence of the quadrupolar modulation, $f(k) = (k/k_0)^n$ with $n=-2$, $-1$, and $1$, where the pivot scale is chosen as $k_0=0.05\hk$.
}
\label{Ap:Table:g*_index}
\end{table*}

\section{Derivations of equations}
\label{Ap:Derivations}
\subsection{Derivation of equation~(\ref{Eq:Integral_constraint})}
\label{Ap:DerivationOfEquation1}

The variance of the mean density perturbation $\bar{\delta}$ (equation~\ref{Eq:mean_density_perturbation}) is given by
\begin{eqnarray}
	  \left\langle \bar{\delta}^2 \right\rangle
	  &=&  \frac{1}{N_{\rm wg}^2} \int d^3r \int d^3x_1\int d^3x_2\,
	  \delta_{\rm D}\left( \vec{r}-\vec{x}_{12} \right) \nonumber \\
	  &\times& \bar{n}(\vec{x}_1)\,\bar{n}(\vec{x}_2)\, \xi(\vec{r},\hat{x}_{1}),
\end{eqnarray}
where we used $\langle \delta(\vec{x}_1)\delta(\vec{x}_2)\rangle = \xi(\vec{x}_{12},\hat{x}_{1})$ with $\vec{x}_{12} = \vec{x}_1-\vec{x}_2$. 
By expanding $\xi(\vec{r},\hat{x}_{1})$ in BipoSHs, we obtain
\begin{eqnarray}
	  \left\langle \bar{\delta}^2 \right\rangle
	  \hspace{-0.25cm}
	  &=& 
	  \hspace{-0.25cm}
	  \frac{1}{N_{\rm wg}^2} \int d^3r \int d^3x_1\int d^3x_2 
	  \delta_{\rm D}\left( \vec{r}-\vec{x}_{12} \right)
	  \bar{n}(\vec{x}_1)\,\bar{n}(\vec{x}_2) \nonumber \\
	  &\times& \left[ \sum_{\ell} \xi_{\ell}(r) {\cal L}_{\ell}(\hat{r}\cdot\hat{x}_{1})
	  + \xi_{\ell\ell'}^{LM}(r) \frac{S_{\ell\ell'}^{LM}(\hat{r},\hat{x}_{1})}{H_{\ell\ell'}^L}\right] \nonumber \\
	  &=&
	  \hspace{-0.25cm}
	  \frac{4\pi}{V} \int dr r^2 \left[ \sum_{\ell} \frac{1}{2\ell+1}Q_{\ell}(r) \xi_{\ell}(r) \right] \nonumber \\
	  &+&
	  \hspace{-0.25cm}
	  \frac{4\pi}{V} \sum_{L\geq 1 M}\sum_{\ell \ell'} \int dr r^2 	
	  \left[   \frac{Q_{\ell \ell'}^{LM*}(r) \xi_{\ell \ell'}^{LM}(r)}{(2\ell+1)(2\ell'+1)(2L+1)(H_{\ell\ell'}^L)^2}  \right] \nonumber \\
\end{eqnarray}
where the survey volume $V$ is estimated as $N_{\rm wg}^2/A$ with $A$ being the normalization factor given by equation~(\ref{Eq:normalization}). We ignore the statistically anisotropic terms $\xi_{\ell\ell'}^{L\geq 1,M}$, because they are clearly subdominant in the integral constraint, resulting in equation~(\ref{Eq:Integral_constraint}).

\subsection{Derivation of equation~(\ref{Eq:full_xi})}
\label{Ap:DerivationOfEquation2}

We start with equation~(\ref{Eq:xiA_average}) and expand the two-point correlation function $\xi(\vec{r},\hat{x}_{1})$ in BipoSHs:
\begin{eqnarray}
	\scalebox{0.84}{$\displaystyle
	\left\langle \widehat{\xi}_{\ell\ell'}^{\ LM}(r)\big|_{\rm A} \right\rangle$}
	\hspace{-0.25cm}&=&
	\hspace{-0.25cm}
	\scalebox{0.84}{$\displaystyle
	\frac{(2\ell+1)(2\ell'+1)(2L+1) H_{\ell\ell'}^L}{A} \int \frac{d^2\hat{r}}{4\pi} \int d^3x_1\, \int d^3x_2 $}\nonumber \\
	&\times&\hspace{-0.25cm} 
	\scalebox{0.84}{$\displaystyle
	\delta_{\rm D}\left( \vec{r} - \vec{x}_{12} \right)\,  S_{\ell\ell'}^{LM*}(\hat{r},\hat{x}_{1})\, \bar{n}(\vec{x}_1)\, \bar{n}(\vec{x}_2) $}\nonumber \\
	&\times&\hspace{-0.25cm}
	\scalebox{0.84}{$\displaystyle
	\left[\sum_{L_1+\ell_1+\ell_1'={\rm even},\, M_1} \xi_{\ell_1\ell'_1}^{L_1M_1}(r)\, \frac{S_{\ell_1\ell'_1}^{L_1M_1}(\hat{r},\hat{x}_{1})}{H_{\ell_1\ell'_1}^{L_1}}  \right] $},
\end{eqnarray}
where we do not consider the integral constraint in the derivation, because the term simply yield $Q_{\ell\ell'}^{LM}(r)\, \langle\bar{\delta}^2\rangle$.
We use the relation
\begin{eqnarray}
	\scalebox{0.93}{$\displaystyle
	S_{\ell\ell'}^{LM*}(\hat{r},\hat{n})
	S_{\ell_1\ell'_1}^{L_1M_1}(\hat{r},\hat{n})$}
	\hspace{-0.25cm}
	&=& 
	\hspace{-0.25cm}
	\scalebox{0.93}{$\displaystyle
	\sum_{L_2M_2}\sum_{\ell_2 \ell'_2}(2L_2+1)(2\ell_2+1)(2\ell'_2+1) $}\nonumber \\
	&\times&
	\hspace{-0.25cm}
	\scalebox{0.93}{$\displaystyle
	T_{\ell\ell';\ell_1\ell'_1;\ell_2\ell'_2}^{LM;L_1M_1;L_2M_2}\, S_{\ell_2\ell_2'}^{L_2M_2 *}(\hat{k},\hat{n}), $}
\end{eqnarray}
where the coefficients $T$ are given by
\begin{eqnarray}
	\hspace{-0.35cm}
	T_{\ell\ell';\ell_1\ell'_1;\ell_2\ell'_2}^{LM;L_1M_1;L_2M_2}
	\hspace{-0.25cm}
	&=& 
	(-1)^{M} (-1)^{\ell+\ell'+L} H_{\ell_1\ell_2}^{\ell} H_{\ell_1'\ell_2'}^{\ell'} \nonumber \\
	&\times& 
	\left( \begin{smallmatrix} L & L_1 & L_2 \\  M & -M_1 & -M_2 \end{smallmatrix}  \right)
	\left\{ \begin{smallmatrix} L & L_1 & L_2 \\  \ell & \ell_1 & \ell_2 \\  \ell' & \ell_1' & \ell_2' \end{smallmatrix}  \right\}.
\end{eqnarray}
We note here that if $\ell+\ell'+L = {\rm even}$ and $\ell_1+\ell'_1+L_1 = {\rm even}$, the coefficients $T$ are zero for $\ell_2+\ell'_2+L_2 = {\rm odd}$. We can then derive $\left\langle \widehat{\xi}_{\ell\ell'}^{\ LM}(r)\big|_{\rm A} \right\rangle$ for $\ell+\ell'+L={\rm even}$ as follows
\begin{eqnarray}
	\left\langle \widehat{\xi}_{\ell\ell'}^{\ LM}(r)\big|_{\rm A} \right\rangle
	 \hspace{-0.25cm}
	&=& 
	\hspace{-0.25cm} \sum_{\ell_1+\ell_1'+L_1={\rm even}}\sum_{\ell_2+\ell_2'+L_2={\rm even}}\sum_{M_1 M_2} \nonumber \\
	&&  (2L+1)(2\ell+1)(2\ell'+1)\, T_{\ell\ell';\ell_1\ell'_1;\ell_2\ell'_2}^{LM;L_1M_1;L_2M_2}
	\nonumber \\
	 &\times& \left( \frac{H_{\ell\ell'}^L}{H_{\ell_1\ell'_1}^{L_1}H_{\ell_2\ell'_2}^{L_2}}  \right) 
	Q_{\ell_2\ell'_2}^{L_2M_2}(r)\, \xi_{\ell_1\ell'_1}^{L_1M_1}(r).
\end{eqnarray}
For $\langle \widehat{\xi}_{20}^{\, 2M}(r)|_{\rm A}\rangle$, we obtain equation~(\ref{Eq:full_xi}).

\section{STANDARD DEVIATION}
\label{Ap:StandardDeviation}

In linear theory, the covariance matrix of $P_{20}^{2M}$ is given by (see equation~(26) in~\citet{Shiraishi:2017wec})
\begin{eqnarray}
	\scalebox{0.85}{$\displaystyle
	{\rm Cov}\left( P_{20}^{2M *}(k),P_{20}^{2M'}(k') \right) \simeq 10\, \delta_{MM'}\, \delta_{kk'}\,\frac{1}{N_{\rm k}} 
	\left( P_0(k) + \frac{1}{\bar{n}_{\rm g}} \right)^2,
	$}
	\label{Ap:Eq:cov}
\end{eqnarray}
where $N_{\rm k}=4\pi k^2 \Delta k V/(2\pi)^3$ is the number of independent Fourier modes in a bin with the survey volume $V$ and the bin width $\Delta k$, $P_0$ is the monopole of the power spectrum, and $\bar{n}_{\rm g}$ is the galaxy mean number density. In the above expression, we ignore higher Legendre multipoles than the monopole because of their smallness. This equation shows that there is no correlation between different two $M$ modes of $P_{20}^{2M}$, and that the covariance does not depend on the value of $M$.

\begin{figure}
	\includegraphics[width=\columnwidth]{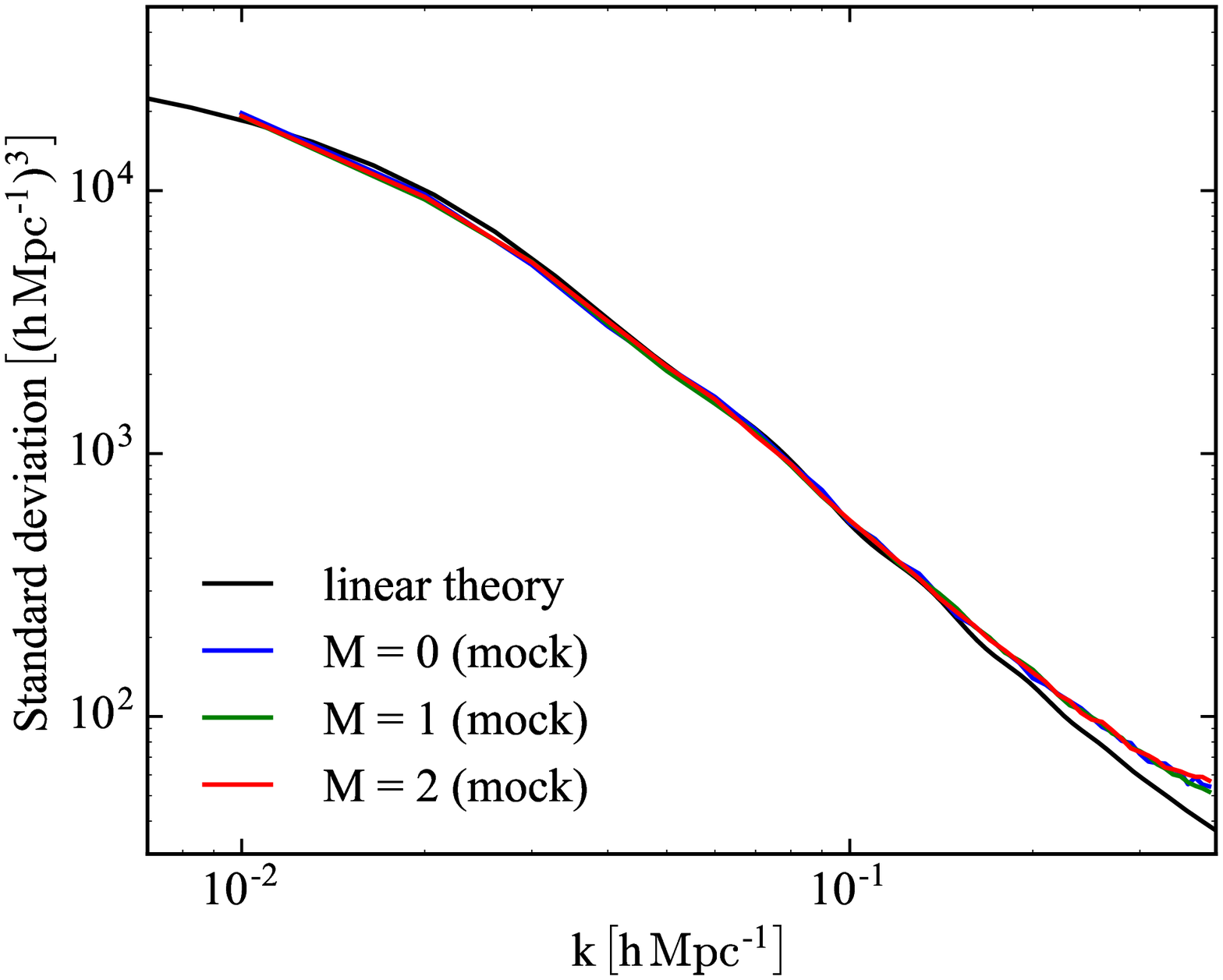}
	\caption{Standard deviations of the BipoSH coefficient $P_{20}^{2M}$ as a function of scales.
	The colored lines represent the estimates of the standard deviation from the QPM mocks (Section~\ref{Sec:CovarianceMatrices}),
	and the black line denotes the prediction of the linear theory given by equation~(\ref{Ap:Eq:cov}).
	}
	\label{Ap:fig:std}
\end{figure}

Figure~\ref{Ap:fig:std} compares the standard deviation of $P_{20}^{2M}$ estimated from the QPM mocks for CMASS NGC in Section~\ref{Sec:CovarianceMatrices} (colored lines) with that computed by equation~(\ref{Ap:Eq:cov}) (black line). We estimate the survey volume and the mean number density for CMASS NGC as $V=2.57\hVGpc$ and $\bar{n}_{\rm g}=2.57\times10^{-4}\hNMpc$, respectively. 
As expected, the standard deviations of $P_{20}^{2M}$ for three $M$ modes, $M=0$, $1$, and $2$, computed from the QPM mocks are closely similar to each other. We find an excellent agreement between the results from the linear theory and the QPM mocks until $k\sim 0.1\hk$, while the linear approximation breaks on smaller scales than $k\sim0.1\hk$.

\bsp	
\label{lastpage}
\end{document}